\documentclass[
	twocolumn,
	amsmath,
	amssymb,
	pra,
	superscriptaddress]
{revtex4}
\usepackage{graphicx}
\usepackage{amsmath,amsfonts,amssymb,amsthm}
\usepackage{fancyhdr}
\usepackage{mathrsfs}
\usepackage{epstopdf}
\usepackage{setspace}
\usepackage{bm}
\usepackage[colorlinks]{hyperref}
\usepackage{siunitx}

\begin{document}

\title{Inelastic scattering of microwave radiation in the dynamical Coulomb blockade}

\author{Juha Lepp\"akangas}
\affiliation{Physikalisches Institut, Karlsruhe Institute of Technology, 76131 Karlsruhe, Germany}

\author{Michael Marthaler}
\affiliation{Institut f\"ur Theorie der Kondensierten Materie, Karlsruhe Institute of Technology, 76131 Karlsruhe, Germany}
\affiliation{ Theoretische Physik, Universit\"at des Saarlandes, 66123 Saarbr\"ucken, Germany}



\begin{abstract}
We study scattering of propagating microwave fields by a DC-voltage biased Josephson junction.
At sub-gap voltages, a small Josephson junction works merely as a non-linear boundary that can absorb,
amplify, and diversely convert propagating microwaves.
In the leading-order perturbation theory of the Josephson coupling energy, the spectral density and quadrature fluctuations
of scattered thermal and coherent radiation can be described in terms of the well-known $P(E)$ function.
Applying this, we study how thermal and coherent radiation is absorbed and amplified in an Ohmic transmission line and in a circuit with a resonance frequency. We show when a coherent input can create a two-mode squeezed output.
In addition, we evaluate scattering amplitudes between arbitrary photon-number (Fock) states, 
characterizing individual photon multiplication and absorption processes occuring at the junction.
\end{abstract}

\maketitle


\section{Introduction}

Charge transport across a mesoscopic constriction is influenced by an interaction with its electromagnetic environment~\cite{Devoret1990,Ingold1992,Holst1994,Hakonen2006,Pashkin2011,Cottet2015,Portier20162}.
In earlier theoretical studies, the electromagnetic degrees of freedom have usually been traced out from the analysis,
with a detailed focus on the mean electric current and its fluctuations.
However, recent microwave-circuit experiments have demonstrated a simultaneous measurement of the current as well as various properties of the emitted and scattered microwave fields~\cite{Hofheinz2011,Reulet2013,Reulet2014,Reulet2015,Portier2014,Rimberg2014,Petta2015,Dima2016,Mottonen2016,Cassidy2017,Fabien2017,Hakonen2017,Jebari2017,Grimm2018,Rolland2018}.
This technological progress has in turn sparked new theoretical efforts to better understand quantum properties of microwave radiation in mesoscopic transport~\cite{Beenakker2001,Marthaler2011,Leppakangas2013,Armour2013,Gramich2013,Belzig2014,Souquet2014,Lambert2015,Hassler2015,Paris2015,Quassemi2015,Dambach2015,Leppakangas2015,Mendes2015,Souquet2016,Portier2016,Leppakangas2016,Dambach2017,Koppenhofer2017,Mendes2018,Simon2018}.
Recent works have predicted and demonstrated effects such as microwave field squeezing~\cite{Armour2013,Reulet2015,Mendes2015,Portier2016,Koppenhofer2017,Mendes2018}, photon-number multistability~\cite{Marthaler2011,Lambert2015},
transmission blockade at certain photon number~\cite{Dambach2015,Souquet2016}, and creation of
anti- or super-bunched microwave photons~\cite{Beenakker2001,Leppakangas2013,Hassler2015,Leppakangas2015,Fabien2017,Grimm2018,Rolland2018}.
In particular, a voltage-biased small Josephson junction has been utilized in applications providing  microwave lasing~\cite{Rimberg2014,Cassidy2017}, parametric amplification~\cite{Jebari2017}, dispersive thermometry~\cite{Dima2016}, and fast single-photon production~\cite{Grimm2018,Rolland2018}.

In this article, we investigate theoretically how different forms of microwave fields scatter by a DC-voltage biased Josephson junction.
We apply an input-output formalism of propagating radiation in a transmission line terminated by a Josephson junction~\cite{Leppakangas2014,Leppakangas2016}.
In the leading-order perturbation theory of the Josephson coupling energy,
average spectral properties of scattered thermal and coherent fields can be described in terms of the $P(E)$ function~\cite{Devoret1990,Ingold1992}.
This function has earlier been used to characterize the average junction current~\cite{Devoret1990,Ingold1992} and current fluctuations (microwave emission)~\cite{Hofheinz2011,Leppakangas2013}.
Here, we study its use in describing microwave scattering,
accounting for an interaction between incoming radiation and junction current.
We derive expressions for microwave absorption, amplification, and quadrature squeezing. 
In addition, we study scattering between individual photon-number states (Fock states) and find
that they are in the same perturbative limit determined by quadrature moments of a continuous-mode displacement operator.

Using the derived expressions, we investigate scattering of different forms of microwave fields in circuits with or without a resonance frequency. 
In particular, we investigate an interplay  between thermal fluctuations and Cooper-pair shot noise in an Ohmic transmission line at low bias voltages.
At higher bias voltages,  we study how different types of radiation get absorbed or amplified~\cite{Jebari2017} when biased close to a resonance condition. 
We also study when a coherent input can create a two-mode squeezed output.
Furthermore, by applying the treatment based on calculating quadrature moments of a continuous-mode displacement operator,
we characterize how incoming arbitrary photon-number in-states get converted into arbitrary photon-number out-states.
The results allow for a straightforward analysis and engineering of arbitrary nonlinear processes
in general electromagnetic environments.

The article is organized as follows. In Sec.~\ref{sec:Model}, we introduce the used input-output formalism and
describe the treatment of various microwave fields as the input: Thermal radiation, coherent radiation, and Fock states.
In Sec.~\ref{sec:MicrowaveConversion1}, we present general formulas describing absorption and conversion of thermal and coherent microwave fields in terms of the $P(E)$ function. We
study absorption and amplification of low- and high-frequency radiation at general bias voltages. 
In Sec.~\ref{sec:Nonclassicality}, we study quadrature fluctuations and squeezing of the created output when having a coherent input.
In Sec.~\ref{sec:MicrowaveConversion2}, we study individual photon multiplication and absorption processes occuring at the junction by evaluating scattering-matrix elements between arbitrary Fock states.
Conclusions and discussion are given in Sec.~\ref{sec:Conclusions}.

\section{The system and the model}\label{sec:Model}

\begin{figure}[t]
\includegraphics[width=\linewidth]{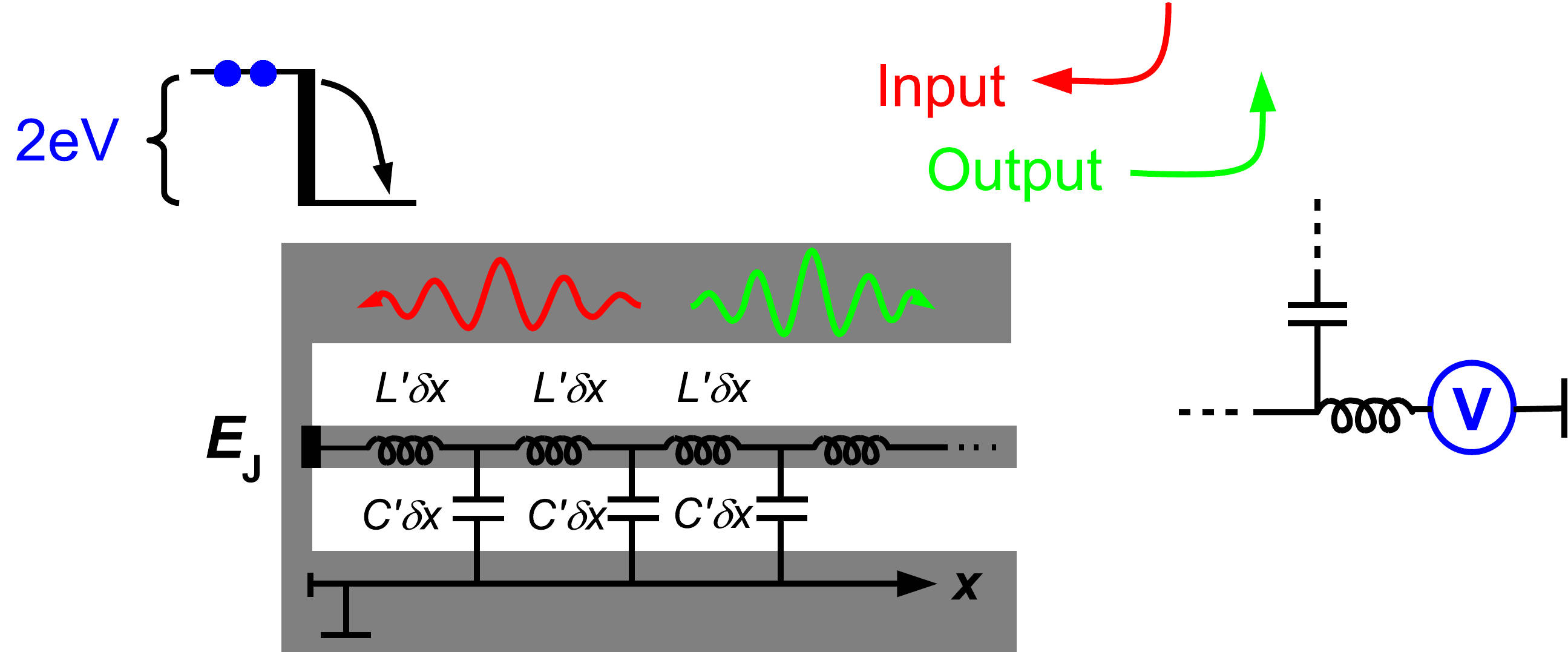}
\caption{
We study scattering of propagating electromagnetic fields in a semi-infinite transmission line terminated by a Josephson junction with coupling energy $E_{\rm J}$.
In the dynamical Coulomb blockade regime, a voltage $V$ is applied across the Josephson junction. Single-Cooper-pair transport is possible via emission of a photon at $2eV/\hbar$, or two (or more) photons at lower frequencies~\cite{Ingold1992}.
Incoming radiation of frequency $\omega$ can interact with Cooper-pair tunneling and, for example, trigger over-bias emission at $2eV/\hbar+\omega$.
}
\label{fig:figure1}
\end{figure}

The microwave circuit we consider is shown in Fig.~\ref{fig:figure1}.
It consists of a DC-voltage biased superconducting transmission line terminated by a Josephson junction with coupling energy $E_{\rm J}$ and critical current $I_{\rm c}=(2e/\hbar)E_{\rm J}$.
In an experimental realization~\cite{Hofheinz2011}, the system can be DC biased via a low-pass filter (inductor) as high-frequency microwave photons propagate and are measured via a high-pass filter (capacitor), see  Fig.~\ref{fig:figure1}.
In the used continuous-mode treatment, the transmission line is described by a semi-infinite lumped-element circuit terminated by the (non-linear) Josephson inductance. We consider mostly explicitly an Ohmic transmission line, i.e., a constant 
inductance $L'$ and capacitance $C'$ per unit length, but
the following analysis can be generalized to setups with resonance frequencies~\cite{Leppakangas2016}.

\subsection{Hamiltonian}
A starting point is the system Hamiltonian~\cite{Leppakangas2016}
\begin{equation}\label{eq:TotalHamiltonian}
H=H_{\rm EE}+H_{\rm J}\, .
\end{equation}
Here $H_{\rm EE}$ describes the electromagnetic environment, in this case the transmission line including the junction capacitor, and $H_{\rm J}$ describes Cooper-pair tunneling across the Josephson junction at the end of the transmission line. The combined Hamiltonian of the junction capacitor and of the semi-infinite transmission line
has the form
\begin{equation}\label{eq:TLHamiltonian}
H_{\rm EE}=\frac{\hat Q^2}{2C}+\sum_{n=1}^{\infty}\left[ \frac{\hat q_n^2}{2C'\delta x} +\frac{1}{2L' \delta x}\left(\hat\Phi_{n}-\hat\Phi_{n-1}\right)^2 \right]\, .
\end{equation}
This is a sum of inductive ($L'$) and capacitive ($C'$) energies per infinitesimal length $\delta x$,
enumerated according to nodes (islands) on the central conductor, and added with the charging energy of the junction capacitor $C$.
The fluxes and charges at each node satisfy $[\hat\Phi_n,\hat q_m]=2ie\delta_{mn}$,
with $\hat \Phi_0$ being the flux and $\hat Q=\hat q_0$ the charge at the Josephson junction. Other combinations of the commutators are zero.
The Cooper-pair tunneling across the Josephson junction is described by the Hamiltonian
\begin{equation}\label{eq:JJHamiltonian}
H_{\rm J}=-E_{\rm J}\cos\left(\omega_{\rm J}t-\frac{2e \hat\Phi_0}{\hbar} \right)\, .
\end{equation}
Here $\omega_{\rm J}=2eV/\hbar$ accounts for the DC-voltage bias.

In the following, we
mark $\hat \phi\equiv (2e/\hbar)\hat \Phi_0$,
which corresponds to the superconducting phase difference across the Josephson junction, relative to the applied DC-voltage $V$.

\subsection{Input-output formalism}
In the limit of vanishing node length, $\delta x\rightarrow 0$,
the Heisenberg equations of motion for the magnetic fluxes at each node convert into a Klein-Gordon wave equation
\begin{equation}
\ddot{\hat\Phi}(x,t)=\frac{1}{L_iC_i}\frac{\partial^2\hat\Phi(x,t)}{\partial^2 x}\, .
\end{equation}
Here $\hat \Phi(x,t)$ is the position-dependent magnetic-flux field operator.
The solution is a propagating electromagnetic field in the transmission line ($x>0$), which we write in the form
\begin{eqnarray}\label{eq:WaveGeneral}
&&\hat \Phi(x,t)=\sqrt{\frac{\hbar Z_{0}}{4\pi}}\int_0^\infty\frac{d\omega}{\sqrt{\omega}}\\
&&\times\left[ \hat a_{\rm  in} (\omega)e^{-i(k_\omega x+\omega t)}+\hat a_{\rm out} (\omega)e^{-i(-k_\omega x+\omega t)}+{\rm H.c.} \right]\, .  \nonumber
\end{eqnarray}
The creation operators of the incoming photons, $\hat a^{\dagger}_{\rm in}(\omega)$, and the corresponding annihilation operators, $\hat a_{\rm in}(\omega)$,
satisfy
\begin{equation}\label{eq:CommutationRelation}
\left[ \hat a_{\rm in}(\omega),\hat a_{\rm in}^{\dagger}(\omega') \right]=\delta(\omega-\omega')\, .
\end{equation}
In a consistent solution, 
this relation is also valid for the operators of the out-field. This has been shown explicitly
to hold at least up to second-order in $E_{\rm J}$ in Ref.~[\onlinecite{Leppakangas2014}].
The wave number $k_{\omega}=\omega\sqrt{C'L'}$ and the characteristic impedance $Z_0=\sqrt{L'/C'}$.

At the Josephson junction ($x=0$) the interaction between the electromagnetic radiation and Cooper-pair tunneling
is described by the boundary condition
\begin{equation}\label{eq:BC}
C\ddot{\hat\Phi}(0,t)-\frac{1}{L'}\frac{\partial\hat\Phi(x,t)}{\partial x}\vert_{x=0}=I_{\rm c}\sin\left[ \omega_{\rm J}t - \hat \phi(t) \right]\, .
\end{equation}
The boundary condition manifests the Kirchhoff's rule for current conservation at the end of the semi-infinite transmission line.

\subsection{Input states}\label{sec:InputStates}
In this article, we consider incoming microwave radiation of the following forms: 
(i) thermal radiation, (ii)  stationary coherent radiation with a thermal background, and (iii) a single- or multi-photon pulse.
Consider first the treatment of a deterministic coherent signal at frequency $\omega_0$.
At zero temperature such state is constructed as~\cite{Loudon}
\begin{eqnarray}
\vert{\rm in}\rangle &=&\hat D \vert 0 \rangle \equiv\vert\alpha\rangle\, .
\end{eqnarray}
Here $\vert 0\rangle$ stands for the continuous-mode vacuum and $\hat D$
is a displacement operator defined as
\begin{eqnarray}
\hat D= \exp\left[\alpha^*\hat a_{\rm in}^\dagger(\omega_0)-\alpha\hat a_{\rm in}(\omega_0)\right]\, .
\end{eqnarray} 
This satifies
\begin{eqnarray}
\hat D^\dagger \hat D=1\, .       \label{eq:UnitaryDisplacement}
\end{eqnarray}
Using the notation $\alpha=\sqrt{2\pi F} e^{i\theta}$, the state gives for the correlation function
\begin{equation}\label{eq:CoherentInputFlux}
\left\langle \hat a^\dagger_{\rm in}(\omega) \hat a_{\rm in}(\omega')\right\rangle=2\pi F \delta(\omega-\omega_0)\delta(\omega'-\omega) \, .
\end{equation}
The photon flux density of the incoming radiation is defined as~\cite{Loudon}
\begin{equation}\label{FluxCoherent}
f_{\rm in}(\omega)\equiv\frac{1}{2\pi}\int d\omega' \left\langle \hat a^\dagger_{\rm in}(\omega) \hat a_{\rm in}(\omega')\right\rangle=F\delta(\omega-\omega_0)\, .
\end{equation}
The total incoming  photon flux is then $F$.

We will now account for thermal radiation at lower frequencies, considering still the case (ii).
In this article, we assume $\omega_0\gg k_{\rm B}T/\hbar$, so that thermal photons do not practically exist at the drive frequency $\omega_0$.
Furthermore, the statistics of thermal radiation is probabilistic: We need to introduce a (formal) density operator $\hat \rho$.
We state that the statistics of incoming radiation is described by
\begin{eqnarray}
\hat \rho=\hat D \hat \rho_{\rm th}\hat D ^\dagger\, ,
\end{eqnarray}
where $\hat \rho_{\rm th}$ describes bare thermal distribution. This state gives for the correlation function
\begin{eqnarray}
\left\langle \hat a^\dagger_{\rm in}(\omega) \hat a_{\rm in}(\omega')\right\rangle &=& \frac{1}{e^{\beta\hbar\omega}-1}\delta(\omega-\omega') \nonumber \\
&+& 2\pi F \delta(\omega-\omega_0)\delta(\omega'-\omega) \, .  \label{eq:ThermalAverage}
\end{eqnarray}
The limit $F=0$ ($\hat D=1$), then corresponds to the case of a bare thermal input, i.e., case~(i).

Finally, in the case of a single-photon input, we have a deterministic input state (pulse)
\begin{eqnarray}\label{eq:SinglePhotonStates}
\vert{\rm in}\rangle=\hat  a_{\xi}^\dagger\vert 0\rangle=\int_0^{\infty} d\omega \xi(\omega) \hat a_{\rm in}^\dagger(\omega)\vert 0\rangle\, .
\end{eqnarray}
Here $\xi(\omega)$ describes the waveform of the incoming photon with normalization $\int_0^\infty d\omega\vert \xi(\omega)\vert^2=1$.
The finite width is needed since any finite photon-number field has a form of a pulse~\cite{Loudon}.
Using Eq.~(\ref{eq:CommutationRelation}) one can show that such single-photon creation and annihilation operators satisfy
\begin{eqnarray}
\left[  \hat a_{\xi}, \hat a_{\xi}^\dagger \right]=1\, .
\end{eqnarray}
Multi-photon states are created analogously. 


\subsection{Solution}
In the input-output theory, the solution of the out field can be expressed as
a function of the time evolution at the boundary~\cite{Book_Gardiner}.
In our problem, the the out field becomes a function of the junction current $I_{\rm J}$~\cite{Leppakangas2016},
\begin{eqnarray}\label{eq:SolutionGeneral}
\hat a_{\rm out}(\omega)&=& \hat a_{0}(\omega) + i \sqrt{\frac{Z_0}{\pi\hbar\omega}} A(\omega) \int_{-\infty}^{\infty}dte^{i\omega t}\nonumber\\
&\times& \hat U^{\dagger}(t,-\infty)\  \hat I_{\rm J}^0(t)\ \hat U(t,-\infty)\, .
\end{eqnarray}
Here the zeroth-order out-operator has the form
\begin{eqnarray}\label{eq:ZerothSolution}
\hat a_{0}(\omega)&=& \frac{ A(\omega)}{ A^*(\omega)}\hat a_{\rm in}(\omega) \, ,
\end{eqnarray}
where the function $A(\omega)$ accounts for the surrounding linear microwave circuit, for an Ohmic transmission having the form
\begin{equation}
A(\omega)=\frac{1}{1-i\omega Z_0 C}\, .
\end{equation}
This function can also incorporate reflections (resonances) in the transmission line~\cite{Leppakangas2016}.
The free-evolution solution  ($E_{\rm J}=0$) for the junction current is
\begin{equation}\label{eq:JunctionCurrent00}
\hat I^0_{\rm J}(t)= I_{\rm c}\sin\left[ \omega_{\rm J}t - \hat \phi_0(t) \right]\, ,
\end{equation}
and the time-evolution is defined by the operator~\cite{Ingold1998}
\begin{equation}\label{eq:TimeEvolution}
\hat U(t,t_0)={\cal T } \exp\left\{ \frac{i}{\hbar}\int_{t_0}^{t} dt' H_{\rm J}^0(t') \right\}\, ,
\end{equation}
where $\cal T$ is the time-ordering operator.
This provides a series expansion of system quantities in powers of the Josephson coupling. 
The phase difference across the Josephson junction has a zeroth-order solution~\cite{Ingold1992,Leppakangas2014}
\begin{equation}\label{eq:PhaseFluctuations}
\hat\phi_0(t)=\frac{\sqrt{4\pi\hbar Z_0}}{\Phi_0}\int_{0}^{\infty}  \frac{d\omega}{\sqrt{\omega}} A(\omega) a_{\rm in}(\omega)e^{-i\omega t}+{\rm H.c.} \, .
\end{equation}
The first- and second-order solutions for the out-field operator $\hat a_{\rm out}(\omega)$ are written down explicitly in Appendix~A.

\subsection{Phase fluctuations}\label{sec:CoherentDriveSolution}

In the evaluation of the spectral density and quadrature fluctuations of the out-field,
a central quantity is the function
\begin{equation}\label{eq:PhaseFluctuations2}
P(t,t')\equiv\left\langle e^{i\hat\phi_0(t)}e^{-i\hat \phi_0(t')} \right\rangle \, .
\end{equation}
This connects Cooper-pair tunneling to the spectral structure of the electromagnetic environment~\cite{Devoret1990,Ingold1992}.
The explicit form of this ensemble average depends on the input field. Let us shortly consider the form of this function in the cases of thermal and coherent inputs.

In thermal equilibrium, the free-evolution phase fluctuations satisfy~\cite{Ingold1992}
\begin{equation}\label{eq:PhaseCorrelatorZeroth1}
P_{\rm th}(t,t')=e^{J_{\rm th}(t-t')}\, , 
\end{equation}
where the phase-correlation function has the form
\begin{equation}\label{eq:PhaseCorrelatorZeroth2}
J_{\rm th}(t) = \left\langle \left[\hat \phi_0(t)-\hat \phi_0(0)\right]\hat \phi_0(0)\right\rangle_{\rm th}\, .
\end{equation}
Using Eqs.~(\ref{eq:ThermalAverage}) and (\ref{eq:PhaseFluctuations}) one obtains
\begin{eqnarray}\label{eq:PhaseCorrelations}
\left\langle\hat \phi_0(t)\hat \phi_0(t')\right\rangle_{\rm th}= 2\int_{-\infty}^{\infty} \frac{d\omega}{\omega}  \frac{ {\rm Re}[Z_{\rm t}(\omega)] }{R_{\rm Q}}\frac{e^{-i\omega(t-t')}}{1-e^{-\beta\hbar\omega}}\, .
\end{eqnarray}
Here $R_{\rm Q}=h/4e^2$ is the superconducting resistance quantum
and we have defined the (real part of the) tunnel impedance as~\cite{Leppakangas2016,Ingold1992}
\begin{equation}\label{eq:TunnelImpedance}
{\rm Re}[Z_{\rm t}(\omega)]=  Z_0\vert A(\omega)\vert^2.
\end{equation}
Note that the quantity $J_{\rm th}(t)$ is a complex valued function, accounting for vacuum fluctuations.


In the case of the coherent-state input we can use the property~\cite{Loudon}
\begin{eqnarray}
\hat D^\dagger \hat a^{(\dagger)}(\omega) \hat D&=&\hat a^{(\dagger)}(\omega)+\alpha^{(*)}\delta(\omega-\omega_0)\, .
\end{eqnarray}
It follows that
\begin{equation}\label{eq:JunctionCurrent0}
 \hat D^\dagger \hat I_{\rm J}^0(t)\hat D =I_{\rm c}\sin\left[ \omega_{\rm J}t - \hat \phi_0(t)-\phi_{\omega_0}(t) \right]\, ,
\end{equation}
where 
\begin{equation}
\phi_{\omega_0}(t)=\frac{\sqrt{4\pi\hbar Z_0}}{\Phi_0} \frac{A(\omega_0)}{\sqrt{\omega_0}}\alpha e^{-i\omega_0 t} + {\rm H.c.}\, .
\end{equation}
Also the time-evolution operator transforms similarly since
\begin{equation}
 \hat D^\dagger H_{\rm J}^0(t)\hat D= -E_{\rm J}\cos\left[ \omega_{\rm J}t-\hat \phi_0(t)-\phi_{\omega_0}(t)\right]\, .
\end{equation}
It follows that the effect of incoming coherent radiation 
can be accounted for by adding an AC-component to the applied voltage $V$. We then get for the relevant phase-fluctuation function
\begin{equation}\label{eq:PhaseFluctuationsCoherent}
P_{\rm coh}(t,t')=e^{i\phi_{\omega_0}(t)}e^{-i\phi_{\omega_0}(t')}e^{J_{\rm th}(t-t')}\, .
\end{equation}
Furthermore, if we assume that $A(\omega_0)\alpha$ is a real number,
the additional phase due to the incoming coherent signal has the form
\begin{equation}\label{eq:Connection0}
\phi_{\omega_0}(t)=a\cos\omega_0 t\, ,
\end{equation}
where $a$ is a real number
\begin{eqnarray}\label{eq:Connection}
a=\sqrt{\frac{8 Z_0}{\omega_0R_{\rm Q}}} \alpha A(\omega_0) \, . 
\end{eqnarray}


\section{Scattering of thermal and coherent radiation}\label{sec:MicrowaveConversion1}
In this section, we investigate scattering of thermal and coherent microwave radiation. 
The results are obtained by a leading-order expansion in the critical current $I_{\rm c}$. The technical calculation involves time integrations of quantities similar to phase-fluctuation function $P(t,t')$, Eq.~(\ref{eq:PhaseFluctuations2}),
and is detailed in Appendix~A.
An energy conservation of the theory is proven in Appendix~B.

We start by presenting general formulas for the photon flux density and power spectral density, describing changes
between the incoming and outgoing microwave fields at certain frequencies. After this, we analyze more detailed specific phenomena predicted by these formulas,
namely conversion and absorption of thermal radiation at low-frequencies and amplification of coherent signals at high frequencies.
Quadrature squeezing of the outgoing field will be studied in Sec.~\ref{sec:Nonclassicality}.
Simultaneous dispersive shift in the reflected field has been investigated theoretically and experimentally in Ref.~\cite{Dima2016}. 

\subsection{General formulas}

The photon flux density of propagating fields in the transmission line is defined as~\cite{Loudon}
\begin{equation}\label{eq:OutputFluxMain}
f_{\rm in/out}(\omega) = \frac{1}{2\pi}\int d\omega'\left\langle \hat a^\dagger_{\rm in/out}(\omega') \hat a_{\rm in/out}(\omega)  \right\rangle\, ,
\end{equation}
and the equivalent power density accounts for the single-photon energy $\hbar\omega$ and is
\begin{equation}
{\cal P}_{\rm in/out}(\omega)=\hbar\omega f_{\rm in/out}(\omega)\, .
\end{equation}

In the following, we present the total result using different contributions in the form
\begin{equation}\label{eq:OutputFluxExplicit}
f_{\rm out}(\omega)- f_{\rm in}(\omega) =  f_{\rm em}(\omega) -f_{\rm abs}(\omega)    \, .  
\end{equation}
The left-hand side then considers the difference between incoming and outgoing photon fluxes
and the right-hand side separates between inelastic contributions where new radiation is created (em)
and incoming radiation is absorbed (abs). 
The term $f_{\rm in}(\omega)$ corresponds to the result for $E_{\rm J}=0$ and is determined by Eq.~(\ref{eq:ThermalAverage}).

\subsubsection{Emission and absorption}

The function $f_{\rm em}(\omega)\geq 0$ describes how radiation is emitted by the junction current fluctuations.
The contribution can be expressed as
\begin{eqnarray}\label{eq:EmissionMain1}
&&f_{\rm em}(\omega)= \frac{1}{1-e^{-\beta\hbar\omega}}\frac{I_{\rm c}^2 {\rm Re}[Z_{\rm t}(\omega)]}{2\omega} \\
&\times&\sum_{\pm}\sum_{n=-\infty}^{\infty}P\left[ \hbar(\pm\omega_{\rm J}+n\omega_0-\omega)  \right] \vert J_n(a)\vert^2   \nonumber \\
&+&\delta(\omega-\omega_0)\frac{I_{\rm c}^2 R_{\rm Q} }{4} \sum_\pm \sum_{n=1}^{\infty} \vert J_n(a)\vert^2  nP\left[\hbar(\pm \omega_{\rm J}-n\omega_0)\right]\, . \nonumber
\end{eqnarray}
This is a function of the probability distribution $P(E)$~\cite{Devoret1990,Ingold1992},
defined as the Fourier transform of the phase correlation function of the thermal field, Eq.~(\ref{eq:PhaseCorrelatorZeroth1}),
\begin{eqnarray}\label{eq:PEDefinition}
P(E)&=&\frac{1}{2\pi\hbar}\int_{-\infty}^{\infty}dt\left\langle e^{i\hat\phi_0(t)}e^{-i\hat\phi_0(0)}\right\rangle_{\rm th} e^{i\frac{E}{\hbar}t}\nonumber \\
&=&\frac{1}{2\pi\hbar}\int_{-\infty}^{\infty}dte^{J_{\rm th}(t)+i\frac{E}{\hbar}t}\, ,
\end{eqnarray}
Here, the plus (minus) sign in front of $\omega_{\rm J}$ corresponds to contribution from  forward (backward) Cooper-pair tunneling.
Since the result for bare thermal radiation corresponds to $a=0$,
we  see that for a finite amplitude $a$, sidebands appear to the emission spectrum, separated by multiples of the drive frequency $\omega_0$.
The summation over $n$ then corresponds to a number of photons exchanged with the drive field in a tunneling process.
The extra contribution  at frequency $\omega_0$ [the bottom line of Eq.~(\ref{eq:EmissionMain1})] describes  emission in the drive mode.

The result for the function $f_{\rm abs}(\omega)\geq 0$ is analogous. It describes how radiation is absorbed by the junction current fluctuations
and has the form
\begin{eqnarray}\label{eq:EmissionMain2}
&&f_{\rm abs}(\omega)= \frac{1}{e^{\beta\hbar\omega}-1}\frac{I_{\rm c}^2 {\rm Re}[Z_{\rm t}(\omega)]}{2\omega} \\
&\times&\sum_{\pm}\sum_{n=-\infty}^{\infty}P\left[ \hbar(\pm\omega_{\rm J}+n\omega_0+\omega)  \right] \vert J_n(a)\vert^2  \nonumber \\
&+&\delta(\omega-\omega_0) \frac{I_{\rm c}^2 R_{\rm Q} }{4} \sum_\pm \sum_{n=1}^{\infty} \vert J_n(a)\vert^2  nP\left[\hbar(\pm \omega_{\rm J}+n\omega_0)\right]\, . \nonumber
\end{eqnarray}
Here, again, the plus (minus) sign in front of $\omega_{\rm J}$ corresponds to contribution from  forward (backward) Cooper-pair tunneling
and the summation over $n$ then corresponds to a number of photons exchanged with the drive field in a tunneling process.
The extra contribution  at frequency $\omega_0$ 
describes the effect of such absorption to the drive mode.

The functions
$f_{\rm em}$ and $ f_{\rm abs}$ are related by the transformation $\omega\rightarrow-\omega$ and $n\rightarrow-n$,
reflecting that emission and absorption are in a quantum theory connected by a sign change in the Fourier transformation of a noise correlation function.

\subsubsection{Junction current fluctuations}
In earlier works, a direct connection between the $P(E)$ theory and junction current fluctuations has been established~\cite{Hofheinz2011,Leppakangas2013,Leppakangas2016}
\begin{eqnarray}
&&\left\langle \hat I_{\rm J}(t)\hat I_{\rm J}(0)\right\rangle_\omega  \label{eq:CurrentFluctuations}      \\
&=&  \pi\hbar   \frac{I_{\rm c}^2 }{2} \sum_{\pm}\sum_{n=-\infty}^{\infty}P\left[ \hbar(\pm\omega_{\rm J}-n\omega_0-\omega)  \right] \vert J_n(a)\vert^2   \, , \nonumber 
\end{eqnarray}
This differs from $ f_{\rm em}$ by not having $\left( 1-e^{-\beta\hbar\omega}\right)^{-1}$ and ${\rm Re}[Z_{\rm t}(\omega)]$ as front factors.
In the zero-temperature limit,
the power density of the emitted radiation is then, up to a frequency-dependent front factor,
the finite-frequency current noise of the Josephson current~\cite{Leppakangas2016}.
At finite temperatures, however, this is not the total spectrum of out-radiation, since
the formula based only on the current fluctuations does not correctly account for changes in the input field.
This becomes evident, for example, when looking the contribution at coherent drive frequencies (delta-function contributions), which are missing in the
current correlator. In an exactly similar way, the
absorption of (and the induced emission to)  the thermal field is not included.
This difference is studied further below.

\subsection{Low-frequency phenomena}\label{sec:EnergyConservation}

\subsubsection{Disappearance of the junction at $V=0$}\label{sec:JunctionDisappearance}
\begin{figure}[tb]
\includegraphics[width=\linewidth]{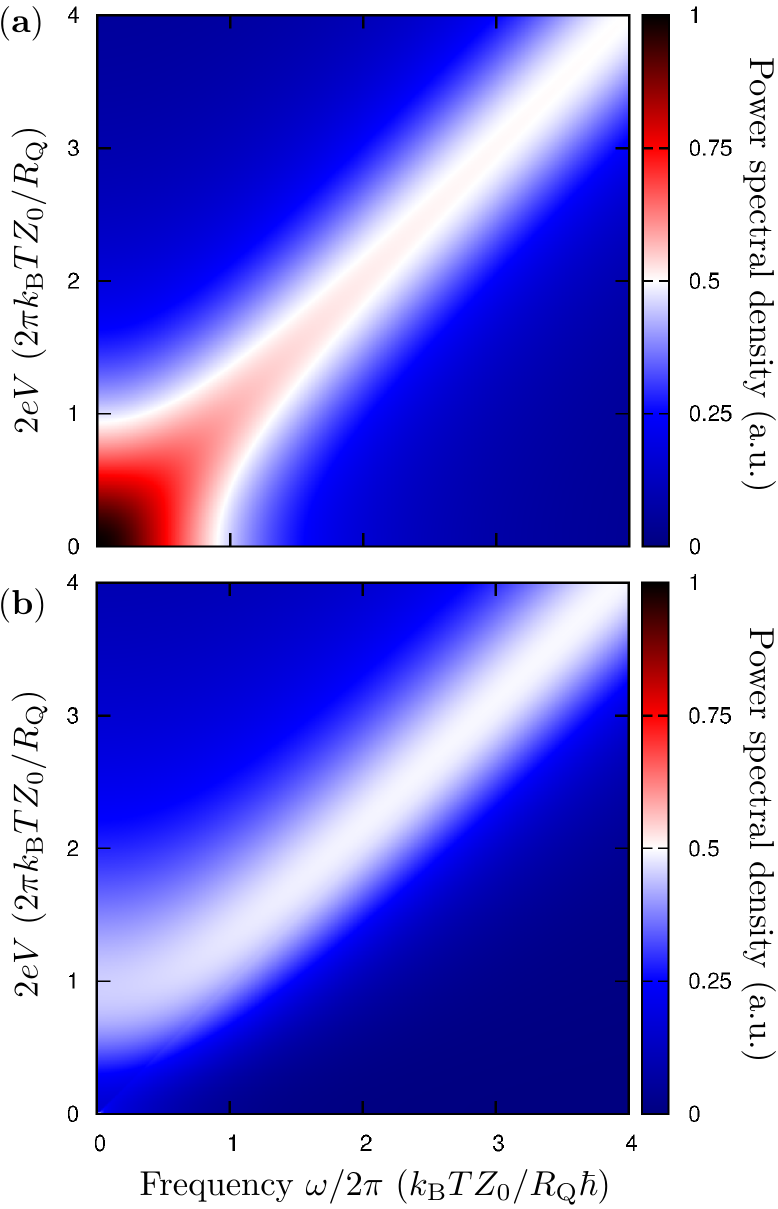}
\caption{
(a) Power spectral density of junction current fluctuations, $\left\langle \hat I_{\rm J}(t)\hat I_{\rm J}(0)\right\rangle_\omega$, of a Josephson junction terminating a voltage $V$ biased Ohmic transmission line with a thermal input.
The current fluctuations occur predominantly at the diagonal line, corresponding to Josephson  frequency $\omega_{\rm J}=2eV/\hbar$.
(b) The simultaneous change in the power density, ${\cal P}_{\rm out}(\omega)-{\cal P}_{\rm in}(\omega)$.
The 'bending' and disappearance of the added power at $2eV/\hbar=0$ manifests that
at zero bias the Josephson junction does not modify the spectrum of the propagating thermal field.
For a finite voltage bias a local maximum of added (zero-frequency) noise appears at $2eV\approx 2\pi k_{\rm B}T Z_0/R_{\rm Q}$.
The parameters are $Z_0=100$~$\Omega$, $ C_{\rm J}=100$~fF, and $T=100$~mK.
}
\label{fig:figure3}
\end{figure}

An illustrative example of changes between incoming and outgoing fields is the case of zero voltage bias, $V=0$,
and the absence of coherent drive, $a=0$.
Here, the power supplied by the voltage source vanishes, $IV=0$, so no changes in the total power is expected.
We can indeed use the detailed balance symmetry of the $P(E)$ function~\cite{Ingold1992}, $P(-E)=e^{-\beta E}P(E)$, to show that
in this case
\begin{eqnarray}\label{eq:NoChangeSpectrum}
 f_{\rm em}(\omega)-f_{\rm abs}(\omega)=0 \, .
\end{eqnarray}
Thus, an undriven and unbiased Josephson junction does not modify the thermal radiation at all:
The outgoing radiation is still thermal and exactly of the same form as the input radiation.
This result is seen in Fig.~\ref{fig:figure3}(b), as the 'avoided crossing' emerging at $2eV/\hbar=0$,
in comparison to the diagonal Josephson AC-current line in Fig.~\ref{fig:figure3}(a).

For higher DC voltages an additional zero-frequency noise appears. For low-Ohmic environments
the noise has a maximum approximately at $2eV= 2\pi k_{\rm B} T R/R_{\rm Q}$, as seen in Fig.~\ref{fig:figure3}(b).
This value corresponds to the width of the current-fluctuation spectrum at $\hbar\omega=2eV$~\cite{Leppakangas2016}.
At higher voltage-biases, the change in the power spectrum approaches the spectrum of the current correlator.
Here, the additional out spectrum becomes a sum of a term proportional to  $\langle \hat I_{\rm J}(t) \hat I_{\rm J}(0)\rangle_{\omega}$ 
and a term
\begin{eqnarray}
&&\frac{1}{e^{\beta\hbar\omega}-1}\frac{\hbar I_{\rm c}^2 {\rm Re}[Z_{\rm t}(\omega)] }{2}\nonumber \\
&&\times \left[ P(2eV-\hbar\omega) -P(2eV+\hbar\omega) \right]\, . \label{eq:Behaviour1}
\end{eqnarray}
In the limit $\omega\rightarrow 0$, this additional contribution becomes proportional to $k_{\rm B}T \partial P(E)/\partial E \vert_{E=2eV}$.
This needs to be compared to $P(2eV)$, the magnitude of  $\langle I_{\rm J}(t)I_{\rm J}(0)\rangle_{\omega=0}$.
For a low-Ohmic transmission line, and in the limit $E\gg k_{\rm B}T$, we have $P(E)\sim Z_0/R_{\rm Q}E$. This means that for $k_{\rm B}T/2eV< 1$
the correction to the low-frequency noise is dominated by the shot noise in the junction current, as also seen in Fig.~\ref{fig:figure3} by the similarity of the two spectral densities at high bias voltages.
Such interplay between transport noise and thermal fluctuations in the power spectrum has been analyzed recently also in Ref.~[\onlinecite{Grabert2016}].

\begin{figure}[tb]
\includegraphics[width=\linewidth]{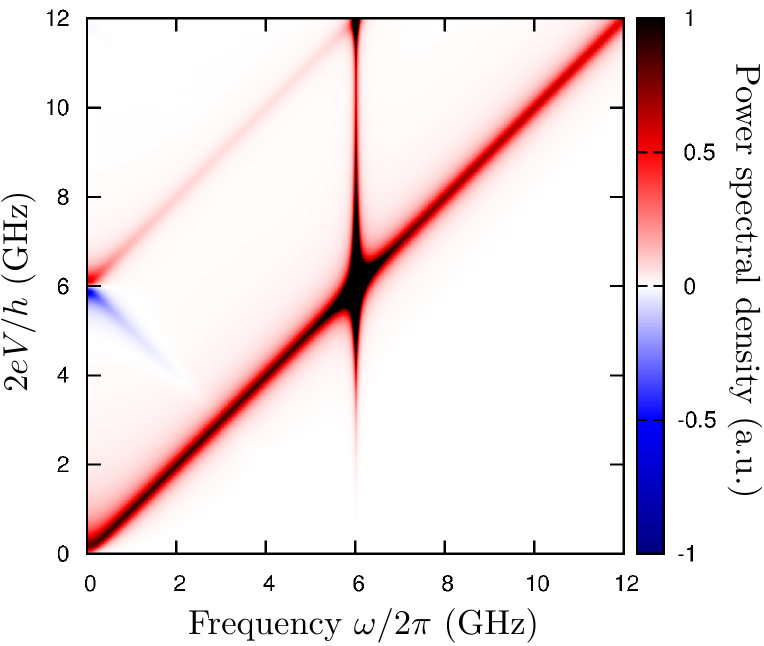}
\caption{Added power, ${\cal P}_{\rm out}(\omega)-{\cal P}_{\rm in}(\omega)$, by a voltage $V$ biased Josephson junction
in a transmission line with a resonance frequency at $\omega_{\rm r}/2\pi=6$~GHz and incoming thermal radiation.
Emission at the Josephson frequency (at the diagonal line) is enhanced close to the resonance frequency $2eV/\hbar=\omega_{\rm r}$.
In particular, when biased just below the resonance frequency, $2eV/\hbar\lesssim \omega_{\rm r}$, emission to the resonance frequency absorbs incoming thermal radiation (blue region at low frequencies). Vice versa, when biased just above the resonance frequency,  emission to the resonance frequency increases low-frequency radiation.
The parameters are $Z_0=100$~$\Omega$,  $ C_{\rm J}=100$~fF, and $T=100$~mK. The impedance ${\rm Re}[Z_{\rm t}(\omega)]$
is added by a Lorentzian corresponding to a $\omega_{\rm r}/2\pi=6$~GHz resonator with quality factor 100 and characteristic impedance $Z_{\rm LC}=100$~$\Omega$.
}
\label{fig:figure4}
\end{figure}

\subsubsection{Low-frequency absorption in a resonance circuit}\label{sec:Heat2}
Absorption of radiation occurs when the contribution $f_{\rm em}(\omega)-f_{\rm abs}(\omega)$  is negative.
Such a result is sound as long as the total flux,
$f_{\rm out}(\omega)=f_{\rm in}(\omega)+f_{\rm em}(\omega)-f_{\rm abs}(\omega)$,   is positive, which
is always possible for small enough coupling energy $E_{\rm J}$.

We obtain that photon-assisted tunneling absorbs thermal energy generally if
\begin{eqnarray}\label{eq:Absorption}
&&P(2eV-\hbar\omega)+ \\
&&\frac{1}{e^{\beta\hbar\omega}-1}\left[ P(2eV-\hbar\omega) -P(2eV+\hbar\omega) \right]<0\, . \nonumber
\end{eqnarray}
Multiplying the left-hand side expression by $\hbar I_{\rm c}^2 {\rm Re}[Z_{\rm t}(\omega)]/2$ one obtains the absorption power density.
Here, the interpretation is that Cooper-pair tunneling extracts thermal radiation as described by the term $P(2eV+\hbar\omega)$, emits
new radiation induced by thermal radiation field, the middle term $P(2eV-\hbar\omega)$, and emits
new radiation induced by vacuum fluctuations, the first term $P(2eV-\hbar\omega)$.

The cooling effect can occur, for example, in a circuit with resonance frequency $\omega_{\rm r}\gg k_{\rm B}T/\hbar$
when voltage biased  below the single-photon emission resonance $\omega_{\rm J}\lesssim \omega_{\rm r}$.
Here, thermal-photon assisted Cooper-pair tunneling, with emission to the resonance frequency, is favored.
This cooling effect is seen as the blue region in Fig.~\ref{fig:figure4}, where we consider a circuit with a resonance frequency at~6~GHz.
This effect works also vice versa: Biasing just above heats the environment.
It is possible to use thermal-photon assisted tunneling to cool electromagnetic degrees of freedom of quantum microwave devices~\cite{Mottonen2016,Silveri2017}.

\begin{figure}[tb]
\includegraphics[width=\linewidth]{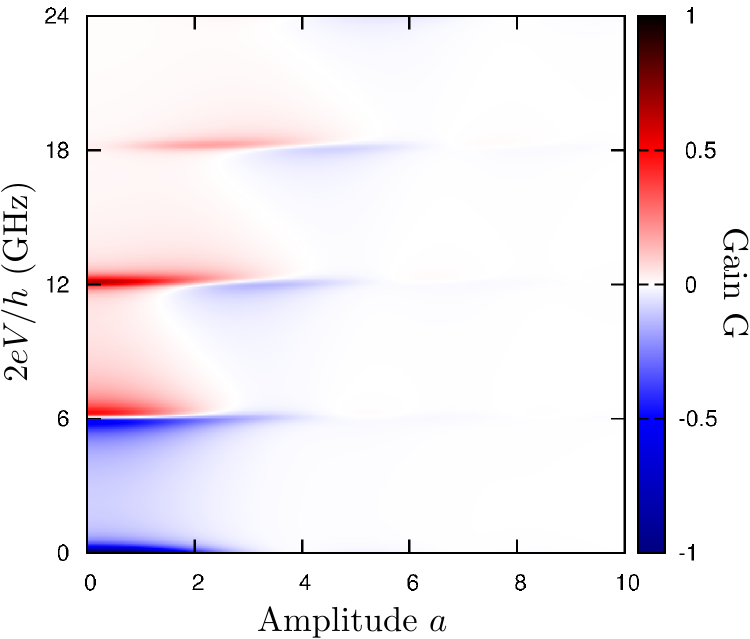}
\caption{
Gain of incoming photon flux for different drive amplitudes $a$ and bias voltages $V$, as defined by Eq.~(\ref{eq:Amplification1}).
We consider a coherent input at $\omega_0/2\pi=6$~GHz and a system as in Fig.~\ref{fig:figure4}.
Depending on the voltage bias, Cooper-pair tunneling can increase or decrease the flux of photons at $\omega_0$.
In particular, just above $2eV/\hbar=\omega_0$ (and $a\lesssim 2$), induced emission to the drive beam occurs through forward Cooper-pair tunneling
and low-frequency photon emission. Just below $2eV/\hbar=\omega_0$, photons are absorbed from the drive beam through backward Cooper-pair tunneling
and again low-frequency photon emission.
The value of $G$ is normalized here by $0.02\pi I_{\rm c}^2{\rm Re}[Z(\omega_0)]/\hbar\omega_0 {\rm GHz}$. 
}
\label{fig:figure5}
\end{figure}

\subsection{Drive field amplification}\label{sec:Heat1}
We study now amplification and absorption of a coherent input at frequency $\omega_0$, i.e.,
flux changes in Eqs.~(\ref{eq:EmissionMain1}) and~(\ref{eq:EmissionMain2}) as given by the delta-function contributions.
This can be characterized by a gain function
\begin{align}\label{eq:Amplification1}
G&= \frac{f_{\rm out}(\omega_0)}{f_{\rm in}(\omega_0)}-1= \frac{4\pi I_{\rm c}^2 {\rm Re}[Z_{\rm t}(\omega_0)]}{\omega_0} \\
&\times \frac{\sum_\pm \sum_{n=-\infty}^{\infty} \vert J_n(a)\vert^2  nP\left[\hbar(\pm \omega_{\rm J}-n\omega_0)\right]}{a^2}  -1. \nonumber
\end{align}
A positive (negative) value of $G$ corresponds to an increase (decrease) of the drive-mode photon flux.
We assume an incoming coherent state $\vert\alpha\rangle$ with photon flux $\vert\alpha\vert^2/2\pi$ and parameter $a$
is obtained by using connection~(\ref{eq:Connection}).

Two competing effects appear when $\omega_{\rm J} \sim \omega_{\rm 0}$.
If the voltage is tuned above the drive, $\omega_{\rm J}\gtrsim \omega_{ 0}$,
induced emission to the drive frequency can occur through forward Cooper-pair tunneling and additional photon emission
to $\omega_{\rm J}-\omega_0$ (or multiphoton emission dissipating the rest energy $\hbar\omega_{\rm J}-\hbar\omega_0$).  
When voltage is tuned below the drive, $\omega_{\rm J}\lesssim \omega_{0}$, drive photons can be absorbed through backward Cooper-pair tunneling and photon emission to $\omega_0-\omega_{\rm J}$ (or multiphoton emission dissipating the rest energy $\hbar\omega_0-\hbar\omega_{\rm J}$). On resonance $\omega_{\rm J}= \omega_{0}$, these two effects cancel each other.
The strength of this effect depends  on the flux density at the drive frequency,
$\sim {\rm Re}[Z_{\rm t}(\omega_0)]/\omega_0$, and on the probability to dissipate the rest energy, $P\left(\hbar\vert\omega_0-\omega_{\rm J}\vert\right)$. 
Similar processes are present also for multi-photon resonances $\omega_{\rm J} \sim n\omega_{\rm r}$, see Fig.~\ref{fig:figure5}.
An experimental realization of this effect is demonstrated in Ref.~\cite{Jebari2017}, achieving near quantum-limited amplification of input signal.

\subsection{Validity region of the perturbative approach and other possible approaches}

In this article, we restrict to a leading-order perturbation theory in powers of the Josephson coupling energy $E_{\rm J}$,
or equivalently the critical current $I_{\rm c}=(2e/\hbar)E_{\rm J}$.
We then implicitly assume the limit of weak interaction, where most of the input radiation is reflected, and
this zeroth-order solution dominates when compared to the higher-order contributions.

A nonperturbative treatment in $E_{\rm J}$ in an electromagnetic environment consisting of two microwave resonators  coupled  weakly to transmission lines is presented in Ref.~[\onlinecite{Leppakangas2018}], where a deterministic photon multiplication is predicted. 
The present article then expands this analysis to account for arbitrary-shaped electromagnetic environments,
with the restriction to the leading-order perturbation theory in the Josephson coupling energy.


\section{Coherent input: Quadrature squeezed output}\label{sec:Nonclassicality}

Microwave radiation can scatter inelastically at a DC-voltage biased Josephson junction.
In this section, we investigate quadrature fluctuations of the created output in such processes.
An electromagnetic field with quadrature fluctuations less than vacuum fluctuations is called quadrature squeezed and
has numerous applications in quantum information and metrology~\cite{WallsMilburn}.

\subsection{Connection to previous works}
Quadrature squeezed microwaves can be created in superconducting transmission lines in the presence of non-linear elements, provided, for example, by  Josephson junctions~\cite{Eichler2011,Nation2012,BJohansson,DCE,Lahteenmaki2013,Schneider2018}.
They can also be emitted by quantum transport:
Inelastic Cooper-pair tunneling produces nonclassical photon pairs
below the classical Josephson radiation peak~\cite{Leppakangas2013,Paris2015,Fabien2017},
and this radiation is ideally quadrature squeezed~\cite{Armour2013}.
However, as being sensitive to junction phase fluctuations,
squeezing with respect to a fixed angle is washed out in a dephasing time that is in typical
experimental conditions less than a microsecond~\cite{Leppakangas2014}, which however can be increased by careful engineering
of the low-frequency impedance.

\begin{figure}[tb]
\includegraphics[width=0.9\linewidth]{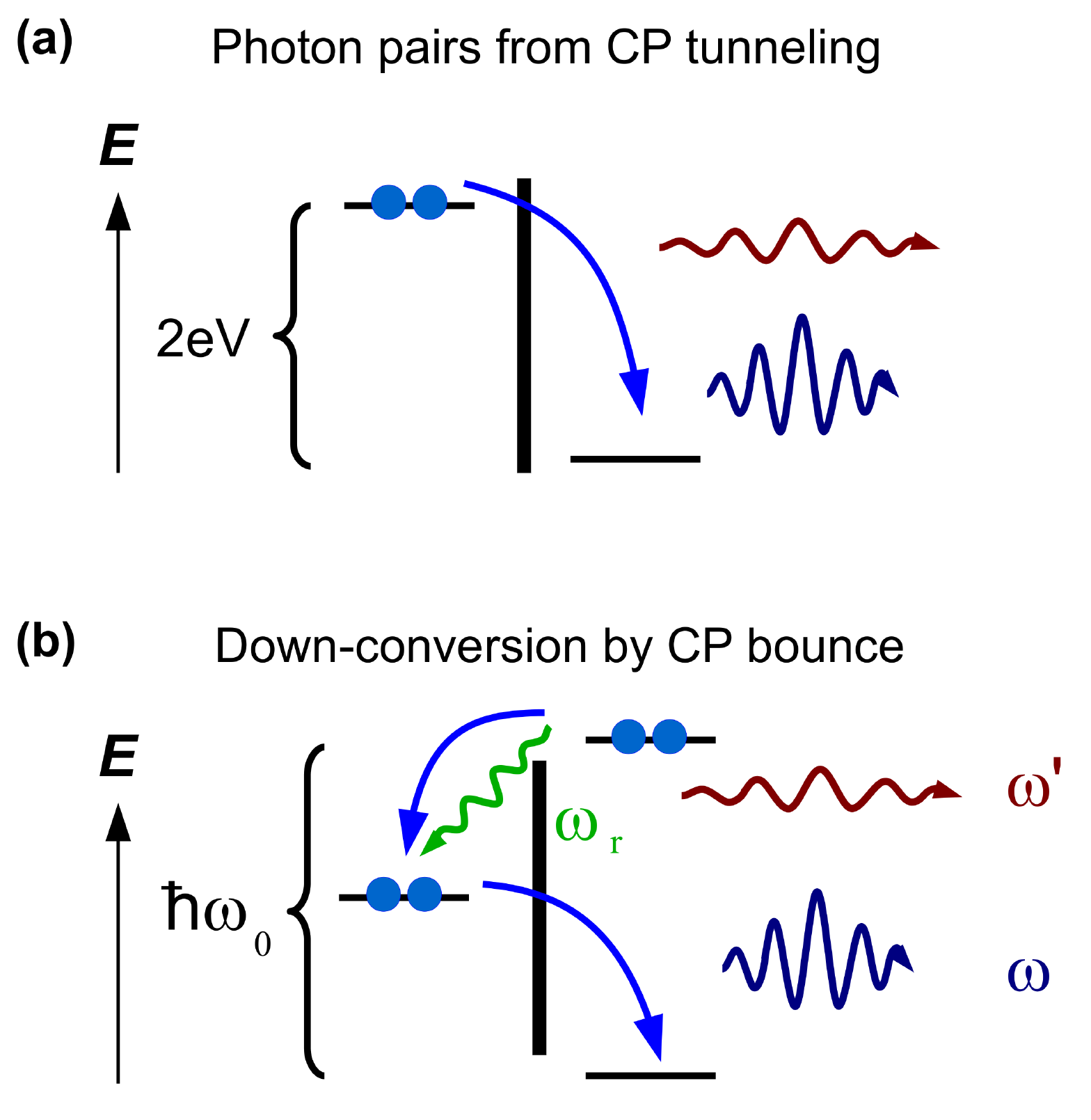}
\caption{
Two ways to produce nonclassical photon pairs and quadrature squeezing by a DC-voltage biased Josephson junction.
In case~(a), two photons are created from the electrostatic energy released in a single Cooper-pair (CP) tunneling event.
The process is similar to a parametric down conversion of a pump-field photon of frequency $\omega_{\rm J}=2eV/\hbar$.
In case~(b), correlated drive-photon assisted backward and forward CP tunneling events produce in total two outgoing
photons, $\omega'$ and $\omega$, whose frequencies sum up to the drive frequency $\omega_0$. In between a virtual photon of frequency $\omega_{\rm r}$ is put (virtually) into the resonator and re-absorbed. 
Only process~(b) is robust against low-frequency voltage fluctuations,
since the total energy of the two outgoing photons does not depend on the fluctuating junction voltage.
A drive-photon assisted single CP tunneling would be equivalent to process~(a) and thereby affected by low-frequency fluctuations.
}
\label{fig:SqueezingProcesses}
\end{figure}

In this section, we find a production of photon pairs and quadrature squeezing by a different process.
This includes
two opposite-direction Cooper-pair tunneling events and splitting of one drive photon of frequency $\omega_0$, Fig.~\ref{fig:SqueezingProcesses}(b).
Such production is thereby robust against low-frequency voltage fluctuations, since the total energy of the two photons does not depend on the bias voltage. 
An analogous phenomenon has been recently predicted~\cite{Portier2016,Mendes2015,Mendes2018} and measured~\cite{Reulet2015} in driven conductors taking use of quasiparticle excitations. 


\subsection{Definition of quadrature squeezing for continuous-mode fields}\label{sec:DefinitionSqueezing}
For this analysis we need a careful definition of quadrature  squeezing for continuous-mode fields~\cite{Loudon}.
We study fluctuations of a field operator
\begin{eqnarray}
\hat O=\int_{\rm BW}d\omega\left[\hat a(\omega) e^{-i\theta(\omega)} + \hat a^\dagger(\omega) e^{+i\theta(\omega)}  \right]\, .
\end{eqnarray}
Here $\rm BW$ stands for the detector bandwidth,
in general introducing a detector filter function~\cite{Miranowicz2010,Leppakangas2016},
in the simplest case just implying integration between two frequencies.
The variable $\theta(\omega)$ corresponds to a continuous-mode generalization of the considered quadrature angle.
Fluctuations of this field are then characterized by the variance
\begin{eqnarray}\label{eq:VarianceDefinition}
{\rm Var}[\hat O]\equiv\left\langle\left( \hat O  - \langle \hat O \rangle\right)^2\right\rangle=\left\langle \hat O^2\right\rangle - \left\langle \hat O\right\rangle^2\, ,
\end{eqnarray}
which can be written as
\begin{eqnarray}
{\rm Var}[\hat O]=\int_{\rm BW}d\omega \int_{\rm BW}d\omega' O(\omega,\omega')\, ,
\end{eqnarray}
where 
\begin{eqnarray}\label{eq:Variance0}
O(\omega,\omega')&=&\delta(\omega-\omega')+2\left\langle \hat a_1^\dagger(\omega) \hat a_1(\omega') \right\rangle +  \nonumber \\
&+&  2\ {\rm Re}\left[  \left\langle \hat a_1(\omega) \hat a_1(\omega') \right\rangle e^{2i\theta(\omega)}  \right]\, .
\end{eqnarray}
The first term on the right-hand side of Eq.~(\ref{eq:Variance0}) is the result for the vacuum, whose total contribution we mark now as
\begin{eqnarray}
{\rm Var}[\hat O]_{\rm vac}=\int_{\rm BW} d\omega   \, . 
\end{eqnarray}
This is also (by definition) the minimum uncertainty of a classical field.

Consider then fluctuations of a squeezed state $\hat D_{\rm sq}\vert 0\rangle$, where
the continuous-mode squeezing operator is defined as~\cite{Loudon}
\begin{eqnarray}
\hat D_{\rm sq}&=&e^{\hat P-\hat P^\dagger} \\
\hat P^\dagger&=&\frac{1}{2} \int_{0}^{\omega_0} d\omega \beta(\omega) \hat{a}^\dagger(\omega)\hat{a}^\dagger(\omega_0-\omega) .
\end{eqnarray}
Here $\beta(\omega)=s(\omega)e^{iv(\omega)}$, with $s(\omega)>0$, accounts for the frequency distribution of the radiation.
It satisfies $\beta(\omega_0-\omega)=\beta(\omega)$.
This creates photon pair states symmetrically around the central frequency $\omega_0/2$.
The corresponding photon flux density is
\begin{equation}\label{FluxSqueezed}
\left\langle \hat a^\dagger(\omega) \hat a(\omega')\right\rangle=\sinh^2\left[s(\omega)\right]\delta(\omega'-\omega)\, .
\end{equation}
The key property of a quadrature squeezed field is a finite value of the correlation function
\begin{eqnarray}\label{FluxSqueezed2}
\left\langle \hat a(\omega) \hat a(\omega')\right\rangle&=&\cosh\left[s(\omega)\right] \sinh\left[s(\omega')\right] e^{iv(\omega')}\\
&\times& \delta(\omega+\omega'-\omega_0)\, . \nonumber
\end{eqnarray}
This leads to the variance
\begin{eqnarray}
&&{\rm Var}[\hat O]={\rm Var}[\hat O]_{\rm vac}+2\int_{\rm BW} d\omega  \sinh^2\left[s(\omega)\right]  \\
&+& 2\int_{\rm BW} d\omega  \cosh\left[s(\omega)\right] \sinh\left[s(\omega)\right] \cos[\theta(\omega)-v(\omega)]\, .  \nonumber
\end{eqnarray}
In particular, the choices $\theta(\omega)=v(\omega)$ and $\theta(\omega)=v(\omega)+\pi$ give the  maximal and minimal variances, respectively,
\begin{equation}
{\rm Var}[\hat O]={\rm Var}[\hat O]_{\rm vac}+\int_{\rm BW} d\omega \left\{ \exp\left[\pm s(\omega)\right]-1 \right\} \,  ,
\end{equation}
where the plus (minus) sign corresponds to the maximum (minimum).
Thus, for $s(\omega)\rightarrow\infty$ and $\theta(\omega)=v(\omega)+\pi$, the total variance approaches zero.

\subsection{Quadrature squeezing in the dynamical Coulomb blockade}

We now redo the calculation of Sec.~\ref{sec:DefinitionSqueezing} but now using the solution obtained
for the scattered field,  Eq.~(\ref{eq:SolutionGeneral}).
We study fluctuations of the (outgoing) transmission-line voltage
\begin{equation}
\hat V=V_0\int_{\rm BW}d\omega\left[\sqrt{\omega}\hat a_{\rm out}(\omega) e^{-i\theta(\omega)} + \sqrt{\omega}\hat a_{\rm out}^\dagger(\omega) e^{+i\theta(\omega)}  \right]\,  ,
\end{equation}
where $V_0=\sqrt{\hbar Z_0/4\pi}$.
The goal is to determine the variance~(\ref{eq:VarianceDefinition}) and compare it to the result for vacuum fluctuations.
When the variance is smaller than the one of vacuum fluctuations, the scattered field is by definition quadrature squeezed.
The vacuum fluctuations are here characterized by  variance 
\begin{eqnarray}
{\rm Var}[\hat V]_{\rm vac}=V_0^2\int_{\rm BW}\omega d\omega \, .
\end{eqnarray}

\subsubsection{Central equations}
We consider now contributions of Eq.~(\ref{eq:Variance0}) beyond the vacuum fluctuations, i.e., beyond the $\delta(\omega-\omega')$ function term.
Let us first evaluate the "diagonal" contribution $\left\langle \hat a_{\rm out}^\dagger(\omega) \hat a_{\rm out}(\omega')\right\rangle$. This corresponds to the photon flux density of the scattered field. In the leading-order perturbation theory in powers of $E_{\rm J}$ this has the form~\cite{Leppakangas2016}
\begin{eqnarray}\label{eq:Variance1}
&&\omega \left\langle \hat a_{\rm out}^\dagger(\omega) \hat a_{\rm out}(\omega') \right\rangle=\pi\delta(\omega-\omega')\\
&&\times\sum_{n\pm}\vert J_{n}(a)\vert^2 I_{\rm c}^2 Z_0\vert A(\omega)\vert^2 P[\hbar(\pm\omega_{\rm J}-n\omega_0-\omega)]\, . \nonumber
\end{eqnarray}
We assume here $\omega,\omega'\gg k_{\rm B}T/\hbar$ and $\omega,\omega'\neq \omega_0$.
This contribution is finite only when $\omega=\omega'$.

The last term is the "non-diagonal" contribution $\left\langle \hat a_{\rm out}(\omega) \hat a_{\rm out}(\omega')\right\rangle$.
Such phase-dependent term is in our system finite only in the presence of a coherent input.
A straightforward calculation gives (Appendix~C)
\begin{align}\label{eq:Variance2}
&\sqrt{\omega\omega'} \left\langle \hat a_{\rm out}(\omega) \hat a_{\rm out}(\omega') \right\rangle= \pi I_{\rm c}^2Z_0A(\omega)A(\omega') \times   \\
&  \sum_{n\pm}\pm iP[\hbar(\omega\mp\omega_{\rm J}+n\omega_0)]  J_{n}(a)J_{n+1}(a)\delta(\omega+\omega'-\omega_0)  \nonumber \, .
\end{align}
Unlike the photon flux density, this contribution can also be negative.
Another important difference is that here the mode frequencies sum up to coherent drive frequency, i.e., $\omega+\omega'=\omega_0$.

Using the simplifying notation
\begin{eqnarray}
\sqrt {Z(\omega)}&\equiv &  \sqrt{Z_0\vert A(\omega)\vert^2}\, , 
\end{eqnarray}
we obtain for the variance of voltage fluctuations of the scattered field 
\begin{widetext}
\begin{eqnarray}\label{eq:SqueezingResult}
&&{\rm Var}[\hat V]-{\rm Var}[\hat V]_{\rm vac} = I_{\rm c}^2V_0^2\int_{\rm BW}d\omega \\
&&\times \left[ \sum_{n\pm}\vert J_n(a)\vert^2 Z(\omega) P[\hbar(\pm\omega_{\rm J}+n\omega_0-\omega)]+ m \sum_{n\pm}J_n(a)J_{n+1}(a)\sqrt{Z(\omega)}\sqrt{Z(\omega_0-\omega)} \left\{ \pm  P[\hbar(\pm\omega_{\rm J}+n\omega_0+\omega)] \right\} \right]\, . \nonumber
\end{eqnarray}
\end{widetext}
We assume here that the integration range includes frequencies symmetrically around $\omega_0/2$
and the introduced variable $m$ is either $-1$ or $1$ (see below).
The left-hand side of Eq.~(\ref{eq:SqueezingResult}) then compares the fluctuations of the out field to vacuum fluctuations.
The integrand of the right-hand side expression then describes quadrature fluctuations between two modes, i.e., two-mode squeezing.
Furthermore,
on the right-hand side we consider only the maximum values of the variance (as a function of squeezing angle $\theta$), which leads to that the introduced variable $m$ is either $-1$ or $1$ (both of them can correspond to the minimum).
In Sec.~\ref{sec:NumericalResults} the value giving a smaller variance is chosen.

In short conclusion,
we find that in the leading-order perturbation theory not only the power spectrum, Sec.~\ref{sec:MicrowaveConversion1}, but also the quadrature fluctuations of the scattered field are described by the $P(E)$-function of the microwave circuit.

\begin{figure}[tb]
\includegraphics[width=0.95\linewidth]{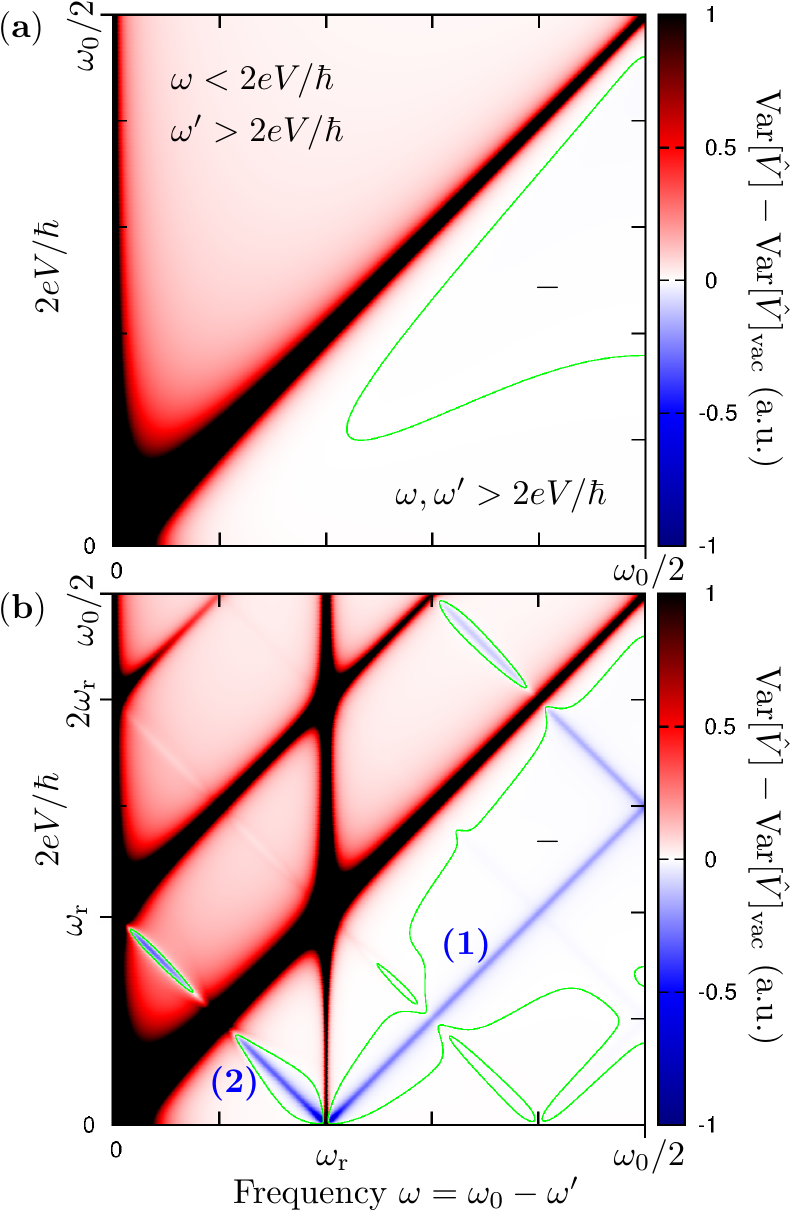}
\caption{
Quadrature fluctuations (two-mode squeezing) of the scattered microwave field when a DC-voltage biased Josephson junction is driven by a coherent field at frequency $\omega_0$.
We show the field variance between modes $\omega$ and $\omega'$ that satisfy $\omega+ \omega'=\omega_0$. 
The (green) contour line corresponds to vacuum fluctuations. The squeezed (negative valued) side of the contour line is indicated by the $-$ sign.
(a) For an Ohmic transmission line,
two-mode squeezing appears in a region where $\omega,\omega'> 2eV/\hbar$. Here emission of photons originating in bare
single-Cooper-pair tunneling is prohibited and emission via the absorption of incoming drive photons dominates.
(b) For a transmission line with a resonance frequency at $\omega_{\rm r}=0.2\omega_0$, two-mode squeezing 
is enhanced at (blues) resonance lines, marked by (1) and (2). At each line, resonant two-photon production
occurs through a down-conversion triggered by a Cooper-pair bounce, see Fig.~\ref{fig:squeezing}(b) and the main text. 
The parameters are as in Fig.~\ref{fig:figure4}, but with  $\omega_0/2\pi=10$~GHz, $\omega_{\rm r}/2\pi=2$~GHz, $Z_{\rm LC}=700$~$\Omega$, and $T=20$~mK. The coherent drive has an amplitude $a=0.75$.
}
\label{fig:squeezing}
\end{figure}

\subsubsection{Numerical results}\label{sec:NumericalResults}



Numerical results for two-mode squeezing created in scattering of incoming
coherent radiation of frequency $\omega_0$ are shown in Fig.~\ref{fig:squeezing}.
We consider scattering in the case of an Ohmic (semi-infinite) transmission line, Fig.~\ref{fig:squeezing}(a),
and a transmission line with a resonance frequency at frequency $\omega_{\rm r}$, Fig.~\ref{fig:squeezing}(b).
We investigate quadrature fluctuations between mode frequencies $\omega$ and $\omega'$ that satisfy $\omega+ \omega'=\omega_0$. 
We plot the scattered-field variance subtracted by the vacuum variance as a function of lower-mode frequency $\omega=\omega_0-\omega'$ and voltage bias (Josephson frequency) $2eV/\hbar$. This means plotting the integrand of Eq.~(\ref{eq:SqueezingResult}) when summed over the values at $\omega$ and $\omega'=\omega_0-\omega$.
Negative values correspond to squeezed regions.

We find that for an Ohmic environment squeezing can occur in a region where $\omega,\omega'> 2eV/\hbar$ [bottom-right corner
of Fig.~\ref{fig:squeezing}(a)]. In this region, most of the radiation is emitted via
drive-photon assisted Cooper-pair tunneling. 
No quadrature squeezing is observed for mode frequencies $\omega<2eV/\hbar$, where emission from direct (not drive-photon-assisted) Cooper-pair tunneling dominates.

In the case of a transmission line with a resonance frequency at $\omega_{\rm r}$,
we find that two-mode squeezing is strongly enhanced near certain resonance lines, labeled by numbers (1) and (2) in Fig.~\ref{fig:squeezing}(b).
The line~(1) corresponds to condition $\hbar\omega=2eV+\hbar\omega_{\rm r}$, or equivalently $\hbar(\omega_0-\omega') = 2eV+\hbar\omega_{\rm r}$ (since $\omega=\omega_0-\omega'$).
At this line,  a two-photon emission, the first one to mode $\omega'$ (open transmission line) and the second one to $\omega_{\rm r}$ (resonator) occurs via a drive-photon assisted backward tunneling, since here (equivalently) $ \hbar\omega_0-2eV=\hbar(\omega'+\omega_{\rm r}) $.
The line labelled as~(1)  corresponds also to a condition for a forward-tunneling process, 
where a photon from the resonator (mode $\omega_{\rm r}$) is absorbed to create a photon to mode $\omega$, i.e., $2eV+\hbar\omega_{\rm r}=\hbar\omega$.
In a higher-order tunneling process, these two opposite-direction Cooper-pair tunnelings occur coherently, with a virtual visit of a photon in the resonator.
In the end, such Cooper-pair bounce produces two propagating photons in the transmission line, one at frequency $\omega$ and another at $\omega'$, which then satisfy $\omega+\omega'=\omega_0$. This bounce process does not suffer from
low-frequency (thermal) voltage fluctuations, since the contribution from the voltage cancels out.
This is the key for having finite phase-dependent (non-diagonal) quadrature correlations and quadrature squeezing.
Another squeezing line seen in Fig.~\ref{fig:squeezing}(b), marked by number (2), corresponds to the condition $\hbar\omega=-2eV+\hbar\omega_{\rm r}$, or equivalently $\hbar(\omega_0-\omega') = -2eV+\hbar\omega_{\rm r}$.
This means the same bounce process as described above, but with reversed Cooper-pair tunneling directions.

We note that the process of Fig.~\ref{fig:SqueezingProcesses}(b) is of fourth order (in $E_{\rm J}$)
in the power spectrum~\cite{Leppakangas2016}. It is then not accounted for by the leading-order theory presented in Sec.~\ref{sec:MicrowaveConversion1}.
It is intriguing to notice that its contribution in the squeezing correlator however appears already in the second order.
This result has been previously obtained and experimentally verified in a similar normal-state system~\cite{Reulet2015}.

\subsubsection{Comparison to previous works}
In comparison to other recent works,
we find that the resonator mode at frequency $\omega_{\rm r}$ plays the same role as virtual population
of quasiparticle states in analogous normal-state systems~\cite{Reulet2015,Portier2016,Mendes2015,Mendes2018}.
Also in our system, squeezing does not emerge when $V=0$.
However, in a very similar setup, quadrature squeezing has been studied and measured in the framework of dynamical Casimir effect~\cite{Nation2012,BJohansson,DCE,Schneider2018}.
Here, one has  $V=0$ but a large Josephson coupling energy $E_{\rm J}$.
In this situation, squeezing of the out field can appear in long-time averages since the down-converted pairs have a fixed relative phase,
provided by the "trapping" of the phase to a local minimum of the Josephson potential energy, $-E_{\rm J}\cos\hat{\phi}$.
This effect then occurs in the limit of large $E_{\rm J}$.


\section{Scattering of Fock states}\label{sec:MicrowaveConversion2}


In this section, we analyze how individual single- and multi-photon states scatter at a DC-voltage biased Josephson junction.
We use here a different theoretical formalism (than $P(E)$ theory):
Rather than evaluating temporal changes in the photon flux density due to Fock-state pulses, a more informative approach here
is the evaluation of the scattering amplitudes
\begin{eqnarray}\label{eq:ResultEntanglement}
T^{1\rightarrow n}&=& T(\omega\rightarrow \omega_1\ldots \omega_n)  \nonumber  \\
&=&\left\langle \left[\hat a_{\rm out}({\omega_1}) \ldots \hat a_{\rm out}({\omega_n}) \right] \hat a^\dagger_{\rm in} (\omega) \right\rangle\, . 
\end{eqnarray}
The ensemble average is made with respect to vacuum state, which we mark as $\vert 0\rangle$.
This gives the amplitude for single incoming photon of frequency $\omega$ converting into $n$ outgoing photons
of frequencies $\omega_1,\ldots, \omega_n$. The total amplitude for an incoming wavepacket is then an integration over individual frequencies with corresponding amplitudes: Integration over $\xi(\omega)$ in Eq.~(\ref{eq:SinglePhotonStates}).
Similarly for multi-photon inputs.

\subsection{Output expressed using a continuous-mode displacement operator}
We start this analysis by making a useful  connection between the leading-order solution for the outgoing field and a continuous-mode displacement operator. 
For this we write the (zeroth-order) phase difference at the junction, Eq.~(\ref{eq:PhaseFluctuations}), in the form
\begin{eqnarray}\label{eq:NewFormPhase1}
i\hat\phi_0(t) &=& i\frac{ \sqrt{4\pi\hbar Z_0} }{\Phi_0}\int_{0}^{\infty}  \frac{d\omega}{\sqrt{\omega}}  A (\omega)\hat  a_{\rm in }(\omega)e^{-i\omega t}-{\rm H.c.}   \nonumber \\
&\equiv& \hat a_{\alpha} - \hat a^\dagger_{\alpha}\, .
\end{eqnarray}
Here, we have defined an unnormalized photon creation operator
\begin{align}\label{eq:NewFormPhase2}
\hat a_{\alpha}^\dagger&=\int_0^\infty d\omega \ \alpha(\omega)\ e^{i\omega t} \ \hat a_{\rm in}^\dagger(\omega) \\
\alpha(\omega) &= -\frac{i}{\Phi_0}\sqrt{\frac{4\pi\hbar Z_0}{\omega}}A^*(\omega)=-i\sqrt{\frac{2Z_0}{R_{\rm Q}\omega}}A^*(\omega)\, .\label{eq:NewFormPhase3}
\end{align}
For simplicity, we now neglect backward Cooper-pair tunneling, i.e., the term $\propto e^{i(\omega+\omega_{\rm J})t}$
in the leading-order solution of operator $\hat a_{\rm out}(\omega)$ (Appendix~A, $a=0$). 
By using the definitions of Eqs.~(\ref{eq:NewFormPhase1}-\ref{eq:NewFormPhase3}), we can then re-express the leading-order solution for the out-field operator  as
\begin{eqnarray}\label{eq:ContinuousModeSolution2}
 \hat a_{\rm out}(\omega) &=&-\frac{\alpha^*(\omega)}{\alpha(\omega)}\hat a_{\rm in}(\omega)  \\
&+& \frac{I_{\rm c}}{4e}\alpha^*(\omega) \int_{-\infty}^{\infty}dt e^{i(\omega-\omega_{\rm J}) t} \exp\left[\hat a_\alpha-\hat a^\dagger_\alpha\right]\, . \nonumber
\end{eqnarray}
The out-field annihilation operator is now defined in terms of a  continuous-mode
displacement operator  
\begin{eqnarray}
\exp\left[\hat a_\alpha-\hat a^\dagger_\alpha\right]\equiv \hat D^\dagger(\alpha) \, .
\end{eqnarray}
This connection originates from the form of the Josephson Hamiltonian ($\cos\hat\phi$ term),
which in the quantized theory makes discrete shifts to the junction charge (Cooper-pair tunneling).

In the following, we exploit several important properties of a continuous-mode coherent state~\cite{Loudon}
\begin{eqnarray}
\vert \alpha(\omega)\rangle \equiv \hat D(\alpha)\vert 0\rangle= \exp\left[ \hat a_{\alpha}^\dagger - \hat a_{\alpha} \right]\vert 0\rangle \, ,
\end{eqnarray}
for which applies
\begin{eqnarray}
\hat a_{\rm in}(\omega)\vert \alpha(\omega) \rangle &=& \alpha(\omega) e^{i\omega t} \vert \alpha (\omega) \rangle \label{eq:SupplementalDProperty1}  \\
\langle 0  \vert \alpha (\omega) \rangle &=& \exp\left[ -\int d\omega \vert \alpha(\omega)\vert^2/2 \right] \label{eq:SupplementalDProperty2} \, .\label{eq:SupplementalDProperty3}
\end{eqnarray}
The operator $\hat D(\alpha)$ itself satisfies
\begin{eqnarray}
\left[ \hat a_{\rm in}(\omega) , \hat D^{(\dagger)}(\alpha)  \right] &=& (-) \alpha(\omega)e^{i\omega t}  \hat D^{(\dagger)}(\alpha)  \, . \label{eq:SupplementalDProperty3}
\end{eqnarray}

\subsection{Evaluation of scattering amplitudes}
Using the connection to the continuous-mode displacement operator, it is straightforward to evaluate general multi-photon scattering amplitudes
to the leading order of critical current $I_{\rm c}$.
We write the general result in the form
\begin{eqnarray}
T^{m\rightarrow n}=T^{m\rightarrow n}_0+T^{m\rightarrow n}_1+\ldots
\end{eqnarray}
according to the order in the critical current $I_{\rm c}$ (or Josephson coupling energy $E_{\rm J}=(\hbar/2e)I_{\rm c}$).

\subsubsection{Single-photon processes}
Let us first consider the zeroth-order contribution $T_0$.
Using Eq.~(\ref{eq:ContinuousModeSolution2}), we get (using $I_{\rm c}=0$)
\begin{equation}
T^{1\rightarrow 1}_0 = -\frac{\alpha^*(\omega_1)}{\alpha(\omega_1)} \left\langle  \hat a_{\rm in}(\omega_1) \hat a^\dagger_{\rm in} (\omega) \right\rangle = -\frac{\alpha^*(\omega_1)}{\alpha(\omega_1)}\delta(\omega-\omega_1)\, .
\end{equation}
Here we used the relation $[\hat a_{\rm in}(\omega),\hat a_{\rm in}^\dagger(\omega')]=\delta(\omega-\omega')$
and that $\hat a_{\rm in}\vert 0\rangle=0$.
This element describes elastic scattering of an incoming photon.
Such contribution exists only in the scattering between single-photon states,
i.e., $T^{1\rightarrow n}_0=0$ for $n>1$.


The leading-order contribution  is obtained by using the term proportional to $I_{\rm c}$ in the out field, giving
\begin{eqnarray}
T^{1\rightarrow 1}_1&=&\frac{I_{\rm c}}{4e} \int_{-\infty}^{\infty}dt e^{i(\omega_1-\omega_{\rm J}) t} \alpha^*(\omega_1)\left\langle  \hat D^\dagger(\alpha)\hat a^\dagger_{\rm in} (\omega)  \right\rangle \nonumber \\
&=& \frac{I_{\rm c}}{4e} e^{-\tilde\rho/2}  \alpha^* (\omega)  \alpha^*(\omega_1) 2\pi\delta (\omega_{\rm J}+\omega-\omega_1)\, , 
\end{eqnarray}
where we have used Eq.~(\ref{eq:SupplementalDProperty1}).
This describes inelastic scattering of an incoming photon, where the electrostatic energy released in a Cooper-pair tunneling
is absorbed,  $\omega_1=\omega+\omega_{\rm J}$.
The appearance of parameter 
\begin{eqnarray}\label{eq:RhoDefinition}
\tilde\rho\equiv \int_{0}^\infty \vert \alpha(\omega)\vert^2=\int_{0}^\infty \frac{d\omega}{\omega} \frac{2{\rm Re}[Z_{\rm t}(\omega)]}{R_{\rm Q}}
\end{eqnarray}
follows from  Eq.~(\ref{eq:SupplementalDProperty2}). [The expression as a function of tunnel impedance
follows from Eqs.~(\ref{eq:NewFormPhase3}) and~(\ref{eq:TunnelImpedance}).]

The parameter $\tilde \rho$  corresponds to the average photon number
of a coherent state $\vert\alpha(\omega)\rangle$,
\begin{eqnarray} 
\tilde \rho = \langle \alpha(\omega)\vert \int d\omega' \hat a^\dagger(\omega') \hat a(\omega') \vert\alpha(\omega)\rangle 
\end{eqnarray}
From Eq.~(\ref{eq:RhoDefinition}) we find that in the case of a finite density of states at $\omega=0$ we have $\tilde\rho\rightarrow\infty$
and consequently the tunneling elements go to zero.
This manifests dephasing due to low-frequency noise. 
As this is a zero-frequency effect,
the amplitudes can be interpreted to have finite momentary values, but vanishing long-time averages.
It can also be shown that the factor $\exp(-\tilde\rho)$ 
is a continuous-mode generalization of the renormalization factor $\exp(-\rho)$ in the $P(E)$-function of a single-mode environment~\cite{Ingold1992}.

\subsubsection{Multi-photon processes}
Let us consider now multi-photon production, i.e., the case $n\geq 2$ in Eq.~(\ref{eq:ResultEntanglement}).
To evaluate this in the first order of $I_{\rm c}$, we
insert the displacement operator once in the $n$ possible out-operators in expression
\begin{eqnarray}
T^{1\rightarrow n} =\left\langle \left[\hat a_{\rm out}({\omega_1}) \ldots \hat a_{\rm out}({\omega_n}) \right] \hat a^\dagger_{\rm in} (\omega) \right\rangle\,   . \nonumber
\end{eqnarray}
We find that the result is non-zero only if the leading-order solution for the out-field is inserted to operators of frequencies $\omega_{n-1}$ or $\omega_{n}$.
We now write the leading-order result in the form
\begin{eqnarray}
T^{1\rightarrow n}_1={\cal T}^{1\rightarrow n}+\sum_{i=1}^n t_i^{1\rightarrow n} \, ,
\end{eqnarray}
where the first part has the form 
\begin{eqnarray}
{\cal T}^{1\rightarrow n}&=& \frac{I_{\rm c}}{2e}\pi e^{-\tilde\rho/2}\delta(\omega+\omega_{\rm J}-\omega_1-\ldots-\omega_n) \nonumber \\   
&\times& \alpha^*(\omega)\alpha^*(\omega_1)\ldots\alpha^*(\omega_n)\, , \label{eq:ScatteringElements}
\end{eqnarray}
which describes $n$-photon emission with energy conservation, $\omega+\omega_{\rm J}=\omega_1+\ldots+\omega_n$:
The electrostatic energy released in a Cooper-pair tunneling is absorbed by the ensemble of photons.
Many different scattering processes are possible, depending on distributions $\alpha(\omega_i)$, but the individual frequencies
always sum up to $\omega+\omega_{\rm J}$.

We also get $n$ additional terms of the form
\begin{eqnarray}\label{eq:SinglePhotonReflection2}
t_1^{1\rightarrow n} &=&  \frac{I_{\rm c}}{2e}\pi e^{-\tilde\rho/2}  \delta(\omega-\omega_1) \delta (\omega_{\rm J}-\omega_2-\omega_3\ldots-\omega_n) \nonumber \\
&\times& \frac{\alpha^*(\omega_1)}{\alpha(\omega_1)} \alpha^*(\omega_2) \alpha^*(\omega_3)\ldots \alpha^*(\omega_n)\, .
\end{eqnarray}
These additional terms manifest an elastic reflection of the input photon of frequency $\omega$ to mode $\omega_i$. The rest of the photons appear due to direct emission, i.e., emission triggered by vacuum fluctuations.
Furthermore, if the background emission is unwanted,
it is possible to reduce it by engineering the tunnel impedance, ${\rm Re}[Z_{\rm t}(\omega)]=Z_0\vert A (\omega)\vert^2\propto \vert \alpha(\omega)\vert^2$, so that it has a small value at frequencies that need to be used in unwanted processes~\cite{Leppakangas2018}.

Consider finally 
scattering amplitudes between general photon-number states.
Assuming an $m$-photon input, we get for the term describing the full conversion to $n$-photon output
\begin{eqnarray}\label{eq:ScatteringElements}
{\cal T}^{m\rightarrow n} &=& \frac{I_{\rm c}}{2e}\pi e^{-\tilde\rho/2} \\   
&\times&\alpha^*(\omega'_1)\ldots \alpha^*(\omega'_m) \alpha^*(\omega_1)\ldots\alpha^*(\omega_n) \nonumber \\
&\times& \delta(\omega'_1+\ldots+\omega'_m+\omega_{\rm J}-\omega_1-\ldots-\omega_n)\, , \nonumber
\end{eqnarray}
where the incoming frequencies are now labeled as $\omega'_1\ldots \omega'_m$. The rest of obtained terms (not listed above)
describe a conversion of only part of the incoming photons, and emission of background photons triggered by vacuum fluctuations, similarly as above.


\section{Conclusions and discussion}\label{sec:Conclusions}
In conclusion, we have analyzed how incoming microwaves of different forms scatter by a DC-voltage biased Josephson junction.
Scattering effects for general circuits can be described in terms of the $P(E)$ function and expectation values of a displacement operator.
The main practical findings were that thermal and coherent radiation 
can be absorbed and amplified in a circuit with a resonance frequency, and that
coherent radiation can be converted into two-mode squeezed microwaves. 
Furthermore,
the non-linear interaction at the junction allows for engineering, in principle, any photon multiplication and multi-photon absorption processes,
with appropriate tailoring of the impedance as seen by the Josephson junction.
Such systems then offer new ways to process quantum microwaves on-chip. 

The analysis presented in this article was made in the perturbative regime,
which provides a simple and intuitive working environment for understanding inelastic Cooper-pair tunneling in general microwave circuits.
In the strict validity region of the perturbative solution, 
the incoming radiation is mostly reflected, rather than scattered to different modes. Alternative
models applicable in the opposite limit~\cite{Armour2013,Gramich2013,Leppakangas2018},
non-perturbative methods through self-consistent expansions~\cite{Marthaler2016}, and
direct calculations of relevant higher-order contributions~\cite{Leppakangas2016,Belzig2014} have also been established.

A future interest is a deeper understanding of quantum information created by quantum transport, and its engineering
by microwave circuit design. In particular, the created radiation can consist of high-photon-number bundles or show very strong photon anti-bunching. The radiation can be tailored to be broadband and fast, or concentrated to narrow resonance frequencies. It can also carry information of the underlying quantum transport. Novel methods and ideas for characterization and detection of quantum information in such forms of microwave light are then of great interest theoretically and experimentally.

\section*{Acknowledgments}
The authors thank M.~Hofheinz, D.~Hazra, A.~Grimm, F.~Blanchet, R.~Albert, S.~Jebari, F.~Portier, A.~Peugeot, I.~Moukharski, P.~Joyez, C.~Altimiras, D.~Vion, B.~Kubala, S.~Dambach, J. Ankerhold, M.~Fogelstr\"om, and G.~Johansson for fruitful discussions at different stages of this project.
This work has been supported by the DFG Grant No.~MA 6334/3-1.


\appendix



\section{Photon flux density}
The evaluation of correlators such as $\left\langle a^{\dagger}(\omega)a(\omega') \right\rangle$  follows guidelines given
in Refs.~\cite{Leppakangas2014,Leppakangas2016}.
The technical difference is the presence of a coherent input at $\omega_0$, inducing sidebands and finite contributions
also for $\omega\neq\omega'$.

\subsection{Flux outside the drive frequency $\omega_0$}
Consider now the evaluation of contribution $\int_0^\infty d\omega'\left\langle \hat a^\dagger_1(\omega) \hat a_1(\omega')\right\rangle$,
where the leading-order solution for the out-field operator $\hat a_{\rm out}(\omega)$ to be inserted here is
\begin{widetext}
\begin{align}
\hat a_1(\omega)&= i I_{\rm c}\sqrt{ \frac{Z_0}{\hbar\omega\pi} } A(\omega)\int_{-\infty}^{\infty}dt e^{i\omega t}\sin\left[ \omega_{\rm J}t +a\cos\omega_0 t -\hat\phi_0(t)\right] \nonumber \\
&= i I_{\rm c}\sqrt{ \frac{Z_0}{\hbar\omega\pi} } A(\omega)\int_{-\infty}^{\infty}dt e^{i\omega t}\left[\frac{1}{2i} e^{i\omega_{\rm J}t-i\hat \phi_0(t)}\sum_{-\infty}^{\infty} i^n J_n(a)e^{i n\omega_0 t} + {\rm H.c.}\right] .
\end{align}
\end{widetext}
This accounts for the incoming coherent field as phase $a(t)=a\cos\omega_0 t$ (we use here a different front sign as in the main text). In the second line we have inserted an expansion in terms of Bessel functions $J_n$,
\begin{equation}
e^{i a\cos\omega_0 t}=\sum_{-\infty}^{\infty} i^n J_n(a)e^{i n\omega_0 t}\, .
\end{equation}
We get for forward Cooper-pair tunneling
\begin{align}
&\left\langle \hat a^{\dagger}_1(\omega) \hat a_1(\omega') \right\rangle  =\int_{-\infty}^{\infty}dt\int_{-\infty}^{\infty}dt' e^{-i(\omega-\omega_{\rm J} )t+i(\omega'-\omega_{\rm J})t'} \nonumber \\
&\times(i)^{n-m}\sum_{n=-\infty}^{\infty}\sum_{m=-\infty}^{\infty}e^{-im\omega_0t'}e^{in\omega_0t}\left\langle e^{-i\hat \phi_0(t)}e^{i\hat \phi_0(t')}\right\rangle \nonumber\\
&\times J_m(a)J_n(a)\, .
\end{align}

In the next step
we do a change of integration variables
\begin{eqnarray}
x=t-t' \,\,\,\,\, \,\,\,\,\, t=\frac{x}{2}+y \nonumber \\
y=\frac{t+t'}{2} \,\, \,\,\,\,\, t'=y-\frac{x}{2} \, .\nonumber
\end{eqnarray}
The integration over the variable $y$ leads to the energy-conservation factor
\begin{equation}
2\pi\delta[ \omega-\omega' - (n-m)\omega_0 ]e^{ix(\omega_{\rm J}-\omega+n\omega_0)}\, .
\end{equation}
We chooce $n=m$, which leads to (including both Cooper-pair tunneling directions)
\begin{align}
&\left\langle \hat a^{\dagger}_1(\omega) \hat a_1(\omega') \right\rangle=\sum_{\pm}\sum_{n=-\infty}^{\infty}\frac{\pi I_{\rm c}^2 Z_0\vert A(\omega)\vert^2}{\omega}\nonumber\\
&\times P\left[ \hbar(\pm\omega_{\rm J}+n\omega_0-\omega)  \right] \vert J_n (a) \vert^2 \delta(\omega-\omega') .   \label{eq:FluxFinal}
\end{align}
This contribution describes emission to frequency $\omega$.

The terms $n\neq m$ give non-diagonal contributions in frequency ($\omega\neq\omega'$). 
These terms describe a beating effect in $f(\omega)= \int d\omega' \left\langle a^{\dagger}(\omega)a(\omega') \right\rangle/2\pi $, 
but do not contribute to long-time averages. 

Additional flux-contributions originating in the second-order solution $\hat a^{\dagger}_2(\omega)$ (given below) appear through terms $\left\langle \hat a^{\dagger}_2(\omega) \hat a_0(\omega') \right\rangle+\left\langle \hat a^{\dagger}_0(\omega) \hat a_2(\omega') \right\rangle$ also appear. When $a=0$,
a direct calculation~\cite{Leppakangas2014} gives
\begin{align}
&\int d\omega'\frac{1}{2\pi}\left[\left\langle \hat a^{\dagger}_2(\omega) \hat a_0(\omega') \right\rangle+\left\langle \hat a^{\dagger}_0(\omega) \hat a_2(\omega') \right\rangle \right] \nonumber\\
&=\frac{1}{e^{\beta\hbar\omega}-1}    \frac{ I_{\rm c}^2 Z_0\vert A(\omega)\vert^2}{2\omega} \nonumber\\
&\times \sum_{\pm }\left[P(\pm \hbar\omega_{\rm J}-\hbar\omega) -  P(\pm\hbar\omega_{\rm J}+\hbar\omega)  \right]\, . \label{eq:NondiagonalZeroA}
\end{align}
The terms proportional to $P(\pm \hbar\omega_{\rm J}-\hbar\omega)$ are  interpreted to describe emission to frequency $\omega$
and the terms proportional to $P(\pm \hbar\omega_{\rm J}+\hbar\omega)$ absorption from frequency $\omega$.

Similarly as above, the presence of a coherent input of amplitude $a$ at frequency $\omega_0$ generalizes this expression to
\begin{align}
&\int d\omega'\frac{1}{2\pi}\left[\left\langle \hat a^{\dagger}_2(\omega) \hat a_0(\omega') \right\rangle+\left\langle \hat a^{\dagger}_0(\omega) \hat a_2(\omega') \right\rangle \right] \nonumber\\
&=\sum_{n=-\infty}^\infty \vert J_n(a)\vert^2\frac{1}{e^{\beta\hbar\omega}-1}    \frac{ I_{\rm c}^2 Z_0\vert A(\omega)\vert^2}{2\omega} \nonumber\\
&\times \sum_{\pm }\left[P(\pm \hbar\omega_{\rm J}+n\hbar\omega_0-\hbar\omega) -  P(\pm\hbar\omega_{\rm J}+n\hbar\omega_0+\hbar\omega)  \right]\, . \label{eq:NondiagonalZeroA}
\end{align}
Again, the terms proportional to $P(\pm \hbar\omega_{\rm J}+n\hbar\omega_0-\hbar\omega)$ are  interpreted to describe emission to frequency $\omega$
and the terms proportional to $P(\pm \hbar\omega_{\rm J}+n\hbar\omega_0+\hbar\omega)$ absorption from frequency $\omega$.
Using the relation
\begin{equation}
1+\frac{1}{e^{\beta\hbar\omega}-1}=\frac{1}{1-e^{-\beta\hbar\omega}}
\end{equation}
we can combine the emission terms with contributions from Eq.~(\ref{eq:FluxFinal}) to get the first contribution on the right-hand side of Eq.~(\ref{eq:EmissionMain1}).

\subsection{Flux at the drive frequency $\omega_0$}
A striking feature of the calculation is the emergence of additional delta-function terms centered at $\omega_0$. These terms describe flux changes of drive photons only, due to photon absorption and induced emission.
We evaluate first
\begin{equation}
\left\langle \hat a^\dagger_0(\omega') \hat a_2(\omega)\right\rangle= \frac{\alpha^*A^*(\omega')}{A(\omega')}\delta(\omega'-\omega_0)\left\langle \hat a_2(\omega)\right\rangle \, .
\end{equation}
Here we have used the fact that an incoming coherent state is a left-hand side eigenstate of operator
$ \hat a_0^\dagger= (A^*/A) \hat a^\dagger_{\rm in}$. After this the ensemble average is subjected to operator
$\hat a_2(\omega)$ only.
We note that if operating with $ \hat a_0^\dagger$ to the thermal background instead (at lower frequencies), we would finally get Eq.~(\ref{eq:NondiagonalZeroA}).
As before, the phase of $\alpha$ is assumed to be the opposite of $A$.

The second-order solution for the out field has the form~\cite{Leppakangas2016}
\begin{widetext}
 \begin{align}
\hat a_2(\omega)  = \frac{E_{\rm J}I_{\rm c}}{\hbar}\sqrt{\frac{Z_0}{ \pi\hbar\omega}}A(\omega)\int_{-\infty}^{\infty}dt\int_{-\infty}^{t}dt'  e^{i\omega t}\left\{  \cos\left[\omega_{\rm J}t'+a\cos\omega_0t' -\phi_0(t') \right]\sin\left[\omega_{\rm J}t+a\cos\omega_0t -\phi_0(t) \right] -{\rm H.c.}  \right\} \, .
\end{align}
\end{widetext}
This again accounts for the incoming coherent field as phase $a(t)=a\cos\omega_0 t$.
The expansion of the sinusoidal functions leads to two terms 
\begin{widetext}
 \begin{align}
 \frac{E_{\rm J}I_{\rm c}}{\hbar}\sqrt{\frac{Z_0}{ \pi\hbar\omega}}A(\omega)\int_{-\infty}^{\infty}dt\int_{-\infty}^{t}dt'  e^{i\omega t}\left[ -\frac{1}{4i}e^{J(t'-t)}e^{i\omega_{\rm J}(t'-t)+a(t')-a(t)}+\frac{1}{4i}e^{J(t'-t)}e^{i\omega_{\rm J}(t-t')+a(t)-a(t')} \right] \, .
\end{align}
and
 \begin{align}
 \frac{E_{\rm J}I_{\rm c}}{\hbar}\sqrt{\frac{Z_0}{ \pi\hbar\omega}}A(\omega)\int_{-\infty}^{\infty}dt\int_{-\infty}^{t}dt'  e^{i\omega t} \left[ -\frac{1}{4i}e^{J(t-t')}e^{i\omega_{\rm J}(t-t')+a(t)-a(t')}+\frac{1}{4i}e^{J(t-t')}e^{i\omega_{\rm J}(t'-t)+a(t')-a(t)} \right] \, .
\end{align}
\end{widetext}

To find frequency correlations that sum to $\omega_0$, the expansion in terms of the Bessel functions have to be
restricted to specific nearby numbers, i.e.,
we expand 
\begin{align}
&e^{-a(t)+a(t')}=\sum_n -i J_{n+1}(a)J_{n}(a)e^{-(n+1)i\omega_0 t}e^{ni\omega_0 t'}\nonumber \\
&=\sum_n -i J_{n+1}(a)J_{n}(a)e^{ni\omega_0 (t'-t)}e^{-i\omega_0 t} \\
&e^{a(t)-a(t')}=\sum_n -i J_{n-1}(a)J_{n}(a)e^{(n-1)i\omega_0 t}e^{-ni\omega_0 t'}\nonumber \\
&=-i J_{n-1}(a)J_{n}(a)e^{ni\omega_0 (t-t')}e^{-i\omega_0 t} \, .
\end{align}
This expansion transforms the two contributions to
\begin{widetext}
 \begin{align}\label{eq:Derivation1}
& \frac{1}{4}\frac{E_{\rm J}I_{\rm c}}{\hbar}\sqrt{\frac{Z_0}{ \pi\hbar\omega}}A(\omega)\int_{-\infty}^{\infty}dt\int_{-\infty}^{t}dt'  e^{i\omega t}  \\
&\times\left[ \sum_n J_{n+1}(a)J_n(a)e^{J(t'-t)}e^{i\omega_{\rm J}(t'-t)+ni\omega_{0}(t'-t)}e^{-i\omega_0t}-\sum_n J_{n-1}(a)J_n(a)e^{J(t'-t)}e^{-i\omega_{\rm J}(t'-t)-ni\omega_{0}(t'-t)}e^{-i\omega_0t} \right] \, . \nonumber
\end{align}
and
 \begin{align}\label{eq:Derivation2}
& \frac{1}{4}\frac{E_{\rm J}I_{\rm c}}{\hbar}\sqrt{\frac{Z_0}{ \pi\hbar\omega}}A(\omega)\int_{-\infty}^{\infty}dt\int_{-\infty}^{t}dt'  e^{i\omega t} \\
&\times\left[ \sum_n J_{n-1}(a)J_n(a)e^{J(-t'+t)}e^{i\omega_{\rm J}(-t'+t)+ni\omega_{0}(-t'+t)}e^{-i\omega_0t}-\sum_n J_{n+1}(a)J_n(a)e^{J(-t'+t)}e^{-i\omega_{\rm J}(-t'+t)-ni\omega_{0}(-t'+t)}e^{-i\omega_0t} \right] \, . \nonumber
\end{align}
\end{widetext}

We do a change of variables $x=t'-t$ and $y=(t'+t)/2$. Unrestricted integration over $y$ can be immediately performed giving $2\pi \delta(\omega-\omega_0)$.
Combining the first term of Eq.~(\ref{eq:Derivation1}) and the second term of Eq.~(\ref{eq:Derivation2}) we get contributions
of type
 \begin{align}
&  \int_{-\infty}^{0}dx  e^{i\bar\omega x} \left[e^{J(x)} -  e^{J(-x)} \right]= 2i  \int_{-\infty}^{0}dx   {\rm Im}[e^{J(x)}]e^{i\bar\omega x} \nonumber\\
&=2i  \int_{-\infty}^{0}dx   {\rm Im}[e^{J(x)}]\left( \cos \bar\omega x +i \sin\bar\omega x  \right)\nonumber\\
&=i  \int_{-\infty}^{\infty}dx   {\rm Im}[e^{J(x)}] {\rm Sgn}(x) \cos \bar\omega x  \nonumber\\
&-  \int_{-\infty}^{\infty}dx   {\rm Im}[e^{J(x)}] \sin\bar\omega x  \, ,
\end{align}
where $\bar\omega=\omega_{\rm J}+n\omega_0$.
In the last form, the first term is purely imaginary, as the other term is purely real.
It will be only the real part that survives in the final expression, after summation also over $\langle  \hat a^\dagger_2(\omega) \hat a_0 \rangle$. This means that such summation contributes as
\begin{align}
&-2\int_{-\infty}^{\infty}dx   {\rm Im}[e^{J(x)}] \sin\bar\omega x \nonumber\\
& =-2\int_{-\infty}^{\infty}dx \frac{1}{2i} \left[ e^{J(x)}-e^{J(-x)} \right]\frac{1}{2i}\left[ e^{i\bar\omega x}-e^{-i\bar\omega x} \right] \nonumber\\
& = 2\pi\hbar \left[ P(\hbar\bar\omega)-P(-\hbar\bar\omega) \right]  \, .
\end{align}
The second term of Eq.~(\ref{eq:Derivation1}) and the first term of Eq.~(\ref{eq:Derivation2}) contribute similarly, with $\bar\omega=-\omega_{\rm J}-n\omega_0$ and an overall minus-sign.

We notice now that all contribitions sum as
\begin{widetext}
 \begin{align}\label{eq:Derivation3}
 2\pi \delta(\omega-\omega_0) \frac{1}{4}\frac{E_{\rm J}I_{\rm c}}{\hbar}\sqrt{\frac{Z_0}{ \pi\hbar\omega}}A(\omega)\left[ \sum_n J_{n+1}(a)J_n(a)+\sum_n J_{n-1}(a)J_n(a) \right] 2\pi\hbar \left[ P(\hbar\omega_{\rm J}+n\hbar\omega_0)-  P(-\hbar\omega_{\rm J}-n\hbar\omega_0) \right] \, .
\end{align}
\end{widetext}
For Bessel functions $J_{n+1}(x)+J_{n-1}(x)=(2n/x)J_n(x)$, which implies
\begin{eqnarray}
J_n(x)J_{n+1}(x)+J_n(x)J_{n-1}(x)=\frac{2n}{x}J_n^2(x) \, .
\end{eqnarray}
This transforms contribution~(\ref{eq:Derivation3}) to
 \begin{align}
& 2\pi \delta(\omega-\omega_0) \frac{1}{4}\frac{E_{\rm J}I_{\rm c}}{\hbar}\sqrt{\frac{Z_0}{ \pi\hbar\omega}}A(\omega) \frac{2n}{a} \nonumber\\
&\times \sum_n \vert J_{n}(a)\vert^2  2\pi\hbar\left[ P(\hbar\omega_{\rm J}+n\hbar\omega_0)-  P(-\hbar\omega_{\rm J}-n\hbar\omega_0) \right] \, .
\end{align}
Multiplying by $\alpha^*A^*(\omega')/[A(\omega')]\delta(\omega'-\omega_0)$,
using the connection $\sqrt{8 Z_0/\omega_0R_{\rm Q}} \alpha A(\omega_0)=-a$,  and integrating over all frequencies,
we get the photon-flux change at the drive mode
\begin{equation}
-\frac{I^2_{\rm c} R_{\rm Q}}{4} \sum_{n=-\infty}^\infty\vert J_{n}(a)\vert^2   \left[ nP[\hbar(n\omega_0+\omega_{\rm J})]+nP[\hbar(n\omega_0-\omega_{\rm J})\right]\, .
\end{equation}

\section{Energy conservation}

The derived equations describe microwave conversion between different frequencies mediated by inelastic Cooper-pair tunneling.
For a consistent theory an energy conservation is crucial.
To show this, we need to compare the power of the propagating electromagnetic radiation to the energy provided  by the voltage source, $IV$.
All quantities have to be calculated to up to the same order in the critical current $I_{\rm c}$.



We consider first the case of thermal input (with coherent amplitude $a=0$).
The key to proving the energy conservation is the relation
\begin{widetext}
\begin{align}\label{eq:EnergyConservation1}
\frac{d}{dt}\exp\left[J(t)\right]=-i \exp\left[J(t)\right]\int_0^\infty d\omega \frac{Z_0 \vert A(\omega)\vert^2}{R_{\rm Q}}  \left[ \left(1+ \coth\left( \frac{1}{2}\beta\hbar\omega \right) \right) e^{-i\omega t} + \left(1-\coth\left( \frac{1}{2}\beta\hbar\omega \right) \right)e^{i\omega t} \right]   \, .
\end{align}
\end{widetext}
A Fourier transformation of this equation gives
\begin{align}\label{eq:EnergyConservation1}
&EP(E)= \\
& \int_{0}^\infty \hbar d\omega \frac{Z_0 \vert A\left(\omega\right)\vert^2}{R_{\rm Q}}P(E-\hbar\omega) \left[1+ \coth\left( \frac{1}{2}\beta\hbar\omega \right)\right]  \nonumber \\
&+ \int_{0}^\infty \hbar d\omega \frac{Z_0 \vert A\left(\omega\right)\vert^2}{R_{\rm Q}}P(E+\hbar\omega) \left[ 1- \coth\left( \frac{1}{2}\beta\hbar\omega \right) \right] \, . \nonumber
\end{align}
Identifying $E=2eV$  and using the leading-order result for the forward Cooper-pair tunneling ($V>0$),
\begin{equation}
I^{+}(V)=\frac{\pi\hbar I_{\rm c}^2}{4e}P(2eV)\, ,
\end{equation}
the work done by the voltage source gets related to the power spectral density,
\begin{align}\label{eq:EnergyConservation2}
& I^+V=  \\
&\frac{I_{\rm c}^2}{4}  \int_{0}^\infty \hbar d\omega Z_0 \vert A\left(\omega\right)\vert^2 P(2eV-\hbar\omega) \left[1+ \coth\left( \frac{1}{2}\beta\hbar\omega \right) \right] \nonumber \\
&+ \frac{I_{\rm c}^2}{4}\int_{0}^\infty \hbar d\omega Z_0 \vert A\left(\omega\right)\vert^2 P(2eV+\hbar\omega) \left[1- \coth\left( \frac{1}{2}\beta\hbar\omega \right) \right]\, . \nonumber
\end{align}
Relation for the backward tunneling is obtained by changing $V\rightarrow -V $. The total current is $I(V)=I^+(V)-I^-(V)=I^+(V)-I^+(-V)$.
Using then relations $\coth(x/2)+1=2e^x/(e^x-1)$, $1-\coth(x/2)=-2/(e^x-1)$, and comparing to Eqs.~(\ref{eq:EmissionMain1}) and~(\ref{eq:EmissionMain2}),  we get
\begin{equation}
IV=\int_0^\infty  d\omega\ \hbar\omega [f_{\rm em}(\omega)-f_{\rm abs}(\omega)]\, .
\end{equation}
This then states that the extra power in the output is the same as the work done by the external voltage source.


For a coherent state input at frequency $\omega_0$ with amplitude  $a>0$ one can proceed similarly.
The Cooper-pair current in this case is generalized to
\begin{equation}
I^{\pm}_{\omega_0}(V)=\sum_{n=-\infty}^\infty \vert J_n(a)\vert^2\frac{\pi\hbar I_{\rm c}^2}{4e}P(\pm 2eV+n\omega_0)\, .
\end{equation}

From Eqs.~(\ref{eq:EmissionMain1}) and~(\ref{eq:EmissionMain2}) we see that
there are two types of contributions in the photon spectrum  $f_{\rm em}(\omega)-f_{\rm abs}(\omega)$:
(i) terms as before but with shifted energy arguments (by $n\hbar\omega_0$) and front factors $\vert J_n\vert^2$, and (ii) terms proportional to delta functions at the drive frequency $\omega_0$.
The generalized junction current $I_{\omega_0}$ has only contributions that are similar to terms (i).

Consider now obtaining the radiation power,  $\int d\omega\ \hbar\omega [f_{\rm em}(\omega)-f_{\rm abs}(\omega)]$,
term by term from the power $I_{\omega_0}V$ by using the $P(E)$-relation given by Eq.~(\ref{eq:EnergyConservation2}) for individual contributions of $I_{\omega_0}V$.
This needs multiplying of Eq.~(\ref{eq:EnergyConservation2}) by $\vert J_n\vert^2$ and changing the energy argument as $E\leftarrow E+n\hbar\omega_0$, meaning
 $V\leftarrow V+n\hbar\omega_0/2e$. Summation over $n$ gives
\begin{widetext}
\begin{align}
& I^+_{\omega_0}V= -\sum_{n=-\infty}^\infty n\frac{\hbar\omega_0}{2e}\vert J_n(a)\vert^2I^+\left(V +n\frac{\hbar\omega_0}{2e}\right) \nonumber \\
&+ \sum_{n=-\infty}^\infty \vert J_n(a)\vert^2\frac{I_{\rm c}^2}{4}  \int_{0}^\infty \hbar d\omega Z_0 \vert A\left(\omega\right)\vert^2 P(2eV+n\hbar\omega_0-\hbar\omega) \left[1+ \coth\left( \frac{1}{2}\beta\hbar\omega \right) \right] \nonumber \\
&+\sum_{n=-\infty}^\infty \vert J_n(a)\vert^2 \frac{I_{\rm c}^2}{4}\int_{0}^\infty \hbar d\omega Z_0 \vert A\left(\omega\right)\vert^2 P(2eV+n\hbar\omega_0+\hbar\omega) \left[1- \coth\left( \frac{1}{2}\beta\hbar\omega \right) \right]\, .
\end{align}
\end{widetext}
On the right-hand side of this equation, the integrations over frequencies match to contributions (i) in the radiation spectrum and the additional contribution proportional to $n\omega_0$ match to terms (ii) in the radiation spectrum.

\section{Squeezing correlator}

We evaluate here the correlator $\left\langle \hat a_{\rm out}(\omega) \hat a_{\rm out}(\omega') \right\rangle$
for frequencies that sum to the drive frequency, $\omega+\omega'=\omega_0$, with $\omega,\omega'>0$.
It is here only the expectation value $\left\langle \hat a_1(\omega) \hat a_1(\omega') \right\rangle$ that gives a finite contribution, i.e., when inserting the leading order solution $\hat a_1$ into operators $\hat a_{\rm out}$.
We get two contributions 
\begin{align}
&c\int_{-\infty}^{\infty}dt\int_{-\infty}^{\infty}dt' e^{i\omega t+i\omega' t'+i\omega_{\rm J}(t-t')} e^{J(t-t')} \nonumber \\
&\sum_{n=-\infty}^{\infty}\sum_{m=-\infty}^{\infty}(i)^n(-i)^m e^{in\omega_0t}e^{-im\omega_0t'} J_n(a)J_m(a)  \, .
\end{align}
and
\begin{align}
&c\int_{-\infty}^{\infty}dt\int_{-\infty}^{\infty}dt' e^{i\omega t+i\omega' t'-i\omega_{\rm J}(t-t')} e^{J(t-t')} \nonumber \\
&\sum_{n=-\infty}^{\infty}\sum_{m=-\infty}^{\infty}(-i)^n(i)^m e^{-in\omega_0t}e^{im\omega_0t'} J_n(a)J_m(a) \, . 
\end{align}
Here $c=-I_{\rm c}^2Z_0A(\omega)A(\omega')/4\sqrt{\omega\omega'}\hbar\pi$.
We do the same change of integration variables as in Appendix~A, integrate over $y$, and obtain the energy-conservation factor for the first term
\begin{equation}
2\pi\delta[\omega+\omega'+(n-m)\omega_0]e^{i(\omega+\omega_{\rm J}+n\omega_0)} \, .
\end{equation}
The energy-conservation factor for the second term has the form
\begin{equation}
2\pi\delta[\omega+\omega'-(n-m)\omega_0]e^{ix(\omega-\omega_{\rm J}-n\omega_0)} \, .
\end{equation}
Assuming $\omega+\omega'=\omega_0$ leads to conditions $n-m=-1$ (the first contribution)
and $n-m=1$ (the second contribution).
Inserting $\int_{-\infty}^\infty dx e^{J(x)+iE/\hbar}=2\pi\hbar P(E)$,
the result is
\begin{widetext}
\begin{align}
\left\langle \hat a_1(\omega) \hat a_1(\omega') \right\rangle= &-c i4\pi^2\hbar\delta(\omega+\omega'-\omega_0)\sum_{n=-\infty}^\infty P[\hbar(\omega+\omega_{\rm J}+n\omega_0)]  J_{n}(a)J_{n+1}(a) \nonumber \\
&-c i4\pi^2\hbar\delta(\omega+\omega'-\omega_0)\sum_{n=-\infty}^\infty P[\hbar(\omega-\omega_{\rm J}-n\omega_0)] J_{n}(a)J_{n-1}(a)\, .
\end{align}
\end{widetext}
Since for odd values of $n$ the Bessel functions satisfy $J_{-n}(a)=-J_n(a)$ and for even values $J_{-n}(a)=J_n(a)$, we have
\begin{align}
&\left\langle \hat a_1(\omega) \hat a_1(\omega') \right\rangle=\sum_{\pm} \mp c i4\pi^2\hbar\delta(\omega+\omega'-\omega_0)\nonumber\\
&\times\sum_{n=-\infty}^\infty P[\hbar(\omega\pm\omega_{\rm J}+n\omega_0)]  J_{n}(a)J_{n+1}(a)\, .
\end{align}

\end{document}